\documentclass[referee,a4paper,12pt,traditabstract]{jswsc} 

\usepackage{graphicx}
\usepackage{txfonts}
\usepackage{subfigure}
\usepackage{epstopdf}
\usepackage[mathlines]{lineno}
\usepackage[authoryear,round]{natbib}
\usepackage[backref]{hyperref}
\usepackage{url}

\hypersetup{colorlinks=true,citecolor=cyan,urlcolor=cyan,linkcolor=blue}

\begin{document}

   \title{First 3D hybrid-Vlasov global simulation of auroral proton precipitation and comparison with satellite observations}

   \titlerunning{Hybrid-Vlasov proton precipitation in 3D}

   \authorrunning{Grandin et al.}

   \author{Maxime Grandin
          \inst{1}\fnmsep\thanks{Corresponding author: \email{\href{mailto:maxime.grandin@helsinki.fi}{maxime.grandin@helsinki.fi}}}
          \and Thijs Luttikhuis\inst{1}
          \and Markus Battarbee\inst{1}
          \and Giulia Cozzani\inst{1}
          \and Hongyang Zhou\inst{1}
          \and Lucile Turc\inst{1}
          \and Yann Pfau-Kempf\inst{1}
          \and Harriet George\inst{1}
          \and Konstantinos Horaites\inst{1}
          \and Evgeny Gordeev\inst{1}
          \and Urs Ganse\inst{1}
          \and Konstantinos Papadakis\inst{1}
          \and Markku Alho\inst{1}
          \and Fasil Tesema\inst{1}
          \and Jonas Suni\inst{1}
          \and Maxime Dubart\inst{1}
          \and Vertti Tarvus\inst{1}
          \and Minna Palmroth\inst{1,2}
          }

   \institute{Department of Physics, University of Helsinki, Helsinki, Finland
         \and
             Space and Earth Observation Centre, Finnish Meteorological Institute, Helsinki, Finland
             }


 
  \abstract
   {
   The precipitation of charged particles from the magnetosphere into the ionosphere is one of the crucial coupling mechanisms between these two regions of geospace and is associated with multiple space weather effects, such as global navigation satellite system signal disruption and geomagnetically induced currents at ground level. While precipitating particle fluxes have been measured by numerous spacecraft missions over the past decades, it often remains difficult to obtain global precipitation patterns with a good time resolution during a substorm. Numerical simulations can help to bridge this gap and improve the understanding of mechanisms leading to particle precipitation at high latitudes through the global view they offer on the near-Earth space system. We present the first results on auroral (0.5--50~keV) proton precipitation within a 3-dimensional simulation of the Vlasiator hybrid-Vlasov model. The run is driven by southward interplanetary magnetic field conditions with constant solar wind parameters. We find that, on the dayside, cusp proton precipitation exhibits the expected energy--latitude dispersion and takes place in the form of successive bursts associated with the transit of flux transfer events formed through dayside magnetopause reconnection. On the nightside, the precipitation takes place within the expected range of geomagnetic latitudes, and it appears clearly that the precipitating particle injection is taking place within a narrow magnetic local time span, associated with fast Earthward plasma flows in the near-Earth magnetotail. Finally, the simulated precipitating fluxes are compared to observations from Defense Meteorological Satellite Program spacecraft during driving conditions similar to those in the simulation and are found to be in good agreement with the measurements.
   }


   \maketitle

\section{Introduction}

One of the crucial challenges in understanding the physics driving space weather lies in the interfaces between regions in the solar--terrestrial domain, from the solar surface to the heliosphere, the magnetosphere, and all the way down to the atmosphere and ionosphere. At those interfaces, numerous and complex coupling mechanisms relate regions consisting of very different plasmas. In particular, magnetosphere--ionosphere (MI) couplings are a particularly active research topic, as they have direct and concrete influences on how space weather affects human-made infrastructures. For instance, particle precipitation leads to enhanced ionisation in the upper atmosphere, increasing the ionospheric conductivities. This facilitates the closure of field-aligned current systems, originating from the magnetosphere, through horizontal currents in the ionosphere. Horizontal currents, in turn, can induce currents at ground level, known as geomagnetically induced currents (GICs). Such GICs pose a threat to power grid networks at high and mid-latitudes \citep[e.g.][]{Clilverd2021,Myllys2014,Rosenqvist2019}, and even at low latitudes \citep{Ebihara2021}, and can flow along railways and pipelines \citep[e.g.][]{Ptitsyna2008,Tsurutani2021}. Besides, precipitating electrons and protons can also lead to the scintillation of signal from navigation satellites \citep[e.g.][]{John2021} and can directly damage the electronics onboard satellites and lead to surface charging \citep[e.g.][]{Ganushkina2021,Horne2013}. Improving our understanding and capability to predict the fluxes of precipitating particles from the magnetosphere into the ionosphere is therefore paramount for devising mitigation strategies to ensure the resilience of human infrastructures in case of a severe geomagnetic event.

While both electrons and protons are known to precipitate from the magnetosphere into the ionosphere, the focus has generally been placed on precipitating electrons, as they provide the majority of the energy input into the ionosphere and are responsible for the most spectacular auroral displays \citep[e.g.][]{Mende2003,Tian2020}. Yet, \cite{Newell2009} estimated that auroral protons can contribute 14--21\% of the energy flux input into the ionosphere, and their contribution can even locally supersede that of electron precipitation \citep[e.g.][]{Hardy1989,Lilensten1998}, for instance in the dusk sector. At energies between 30 and 80~keV, the proton energy flux was also found to be higher than the electron energy flux by nearly one order of magnitude during geomagnetic storms \citep{Tian2020}. Additionally, precipitating protons are more efficient at producing ionisation and at increasing the value of the Pedersen conductivity than precipitating electrons \citep{Galand2004}. Proton precipitation therefore needs to be taken into account when making estimates of the energy budget associated with particle precipitation from the magnetosphere into the ionosphere. 

Another reason for studying precipitating protons is that, contrary to auroral electrons which are significantly accelerated by electric potential drops above the ionosphere, protons can be used fairly reliably to map effects observed in the ionosphere all the way to the magnetotail. Hence, they provide a valuable means to study the complex and elusive tail processes by using ground-based instrumentation \citep{Liang2013}. A recent review paper by \cite{GallardoLacourt2021} provides an overview of the current understanding of auroral proton precipitation and its optical signatures at subauroral latitudes, as well as their links with and differences from other optical phenomena observed equatorwards from the main auroral oval.

There are various mechanisms that can lead to the precipitation of protons with energies on the order of hundreds of electronvolts to tens of kiloelectronvolts. First, when the magnetic field topology changes through magnetic reconnection at the dayside magnetopause (or in the magnetotail), protons can be injected on a newly open (or newly closed) field line and, if their pitch angle allows, they can follow it all the way down to the upper atmosphere. This mechanism can produce nightside proton aurora during substorm injections \citep[e.g.][]{Soraas2018} as well as cusp aurora on the dayside \citep[e.g.][]{Bryant2013}. Second, pitch-angle scattering of protons can take place on highly stretched magnetic field lines in the magnetotail due to the field line curvature radius becoming comparable to the proton gyroradius \citep{Sergeev1982,Sergeev1983}. This is believed to be the main mechanism to produce the diffuse proton aurora at nightside magnetic local times \citep[e.g.][]{Donovan2003}. Third, protons can be scattered into their loss cone through interactions with electromagnetic ion cyclotron (EMIC) waves \citep[e.g.][]{Cao2016,Jordanova2007,Liang2014}. Such waves are often generated in the prenoon and evening sectors of the inner magnetosphere \citep{Saikin2015,Usanova2012}, with a tendency to become stronger in the dusk sector during active geomagnetic conditions \citep{Yue2019}. Recent studies showed that this mechanism is the dominant source for proton precipitation at subauroral latitudes \citep{Shreedevi2021,Zhu2021}. A proton cloud injected into the inner magnetosphere during a substorm can also provide suitable conditions for the growth of EMIC waves which lead to the precipitation of those protons as they drift towards the dusk sector \citep{Yahnin2021}.

Observations of proton precipitation are available both from ground-based optical instruments \citep[see][for a review on the topic]{Galand2006} and from various satellites. The latter can either provide direct measurements of particle fluxes or use remote sensing methods that enable their estimation \citep[e.g.][]{Galand2004,Knight2012}. Nevertheless, observations alone provide only a limited spatial coverage of near-Earth space, and suitable instrument conjunctions enabling one to relate ionospheric signatures to magnetospheric processes are rare. 

For this reason, numerical simulations are a valuable tool to bridge the gap between observations and to provide an understanding of the couplings between regions of near-Earth space. There exist, for instance, kinetic transport models that are able to calculate auroral emission altitude profiles resulting from given incoming precipitating particle fluxes, such as the TRANS4 model \citep{Simon2007}. Such codes can then be coupled to global magnetospheric models or MI coupling models that provide precipitating particle fluxes as an input \citep[e.g.][]{Maggiolo2012}. 

Global magnetospheric models, in turn, can shed light on the sources of precipitating particles. Magnetohydrodynamics (MHD) simulations can estimate the integrated energy input of electron precipitation through empirical formulae based on magnetospheric quantities \citep[e.g.][]{Palmroth2006}, but this does not give access to the energy spectra of precipitating particles. A possible solution to go beyond this description is to inject test particles in the global MHD simulation. For instance, \cite{Connor2015} carried out three-dimensional (3D) simulations coupling the Open Global Geospace Circulation Model with the Liouville Theorem Particle Tracer code and studied the influence of the IMF clock angle on the energy dispersion patterns for precipitating protons in the cusp. Their simulations reproduced observations and linked cusp ion signatures to the type (antiparallel or component) and location site (subsolar point, lobes, flanks) of magnetic reconnection on the dayside. 

To account for processes taking place at kinetic scales, beyond-MHD descriptions are needed. Using 2.5D hybrid simulations (in which the ions evolve in a 2D space domain but electromagnetic fields have components in 3D; electrons are treated as a fluid), \cite{Omidi2007} related the properties of poleward moving auroral forms observed in ground-based optical data to those of flux transfer events (FTEs) transiting in the cusp. The work of \cite{Omidi2007} is an example of the hybrid particle-in-cell (PIC) approach, in which ion velocity distributions are constructed through statistics over a large number of macroparticles evolving kinetically. 

An alternative hybrid-kinetic approach is the hybrid-Vlasov description, in which ion velocity distributions are discretised on a velocity grid and propagated using e.g. a semi-Lagrangian or a Eulerian scheme. A prime example of the hybrid-Vlasov approach is the Vlasiator model \citep{Palmroth2018}. The development of the Vlasiator code, which started in 2009, has been carried out always targeting upcoming resources and future improvements in computational capability rather than already existing ones, enabling the model to be run on the most powerful supercomputers at a given time \citep[see][for details on how Vlasiator came to be]{Palmroth2022}. Vlasiator has been used to study foreshock waves \citep[e.g.][]{Turc2019}, bow shock reformation \citep{Johlander2022}, flux transfer events at the dayside magnetopause \citep{Hoilijoki2019}, magnetosheath jets \citep{Suni2021}, radial diffusion in the outer radiation belt \citep{George2022}, as well as magnetotail ion distributions \citep{Runov2020}, among many other topics. Two past studies made use of Vlasiator to investigate mechanisms leading to nightside and dayside proton precipitation in 2D setups \citep{Grandin2019,Grandin2020}. In this paper, we present for the first time self-consistent proton precipitation fluxes obtained in a global 3D hybrid-Vlasov simulation performed with Vlasiator.

This paper is organised as follows: in Section~\ref{sec:data_model}, we provide a brief overview of the Vlasiator model and of the 3D run used in this study, followed, in Section~\ref{sec:mapping}, by a description of the method used to map precipitating proton fluxes calculated near the inner boundary of the Vlasiator simulation domain (i.e. in the inner magnetosphere) to ionospheric altitudes. Section~\ref{sec:results} then presents the results, starting with a description of the global proton precipitation patterns throughout the last 600~s of the Vlasiator run, then analysing local proton energy spectra in the cusp and on the nightside, and investigating geomagnetic latitude (MLAT) and magnetic local time (MLT) variations of the precipitating proton fluxes. In the last subsection presenting the results, the precipitating proton fluxes obtained with Vlasiator are compared to observed fluxes from the Special Sensor J (SSJ) instrument on Defense Meteorological Special Program (DMSP) spacecraft. Section~\ref{sec:discussion} then provides a discussion of the results, and the conclusions of the study are summarised in Section~\ref{sec:conclusions}.

\section{Data and Model}
\label{sec:data_model}

\subsection{Vlasiator simulation}

Vlasiator \citep{Palmroth2018,Vlasiator_code} is a global hybrid-Vlasov model of near-Earth space plasma. In the hybrid-Vlasov description, ions (protons only in the runs discussed here) are described through their velocity distribution function (VDF) that is evolved in time with the Vlasov equation, whereas electrons are treated as a massless, charge-neutralising fluid. The electromagnetic fields are propagated in time with Maxwell's equations in the Darwin approximation \citep[i.e. the displacement current term is neglected in Maxwell--Amp\`ere's equation;][]{Kaufman1971}, and the system is closed with a modified Ohm's law including the Hall term and a polytropic description of the electron pressure gradient term. The simulation domain is six-dimensional (6D, or ``3D--3V''): the three-dimensional ordinary space is discretised on a Cartesian grid in the Geocentric Solar Ecliptic (GSE) frame, and in each ordinary-space cell the ion VDF is discretised on a 3D Cartesian velocity grid. The ordinary-space grid uses adaptive mesh refinement (AMR) to improve the spatial resolution in regions where processes operate over finer scales (tail current sheet, magnetopause, magnetosheath, bow shock); details on the spatial AMR can be found in \cite{Ganse2023}. The Earth's magnetic field is modelled as an ideal dipole field whose axis is aligned with the $Z_\mathrm{GSE}$ direction.

In this paper, we make use of a 3D--3V run in a simulation domain extending from $X = -110\,R_\mathrm{E}$ in the magnetotail to $X = 50\,R_\mathrm{E}$ in the solar wind and confined within $|Y| < 58\, R_\mathrm{E}$ and \mbox{$|Z| < 58\, R_\mathrm{E}$}. The inner boundary is a sphere centred at the Earth and with a radius of $4.7\,R_\mathrm{E}$ behaving like a perfect conductor and where VDFs are fixed Maxwellian distributions. The ordinary-space AMR grid has three refinement levels in addition to the base grid of 8000\,km resolution, yielding a finest resolution of 1000\,km in the tail current sheet and at the dayside magnetopause. The velocity space resolution is 40\,km~s$^{-1}$, and the simulation uses a sparsity threshold of $10^{-15}$\,m$^{-6}$~s$^3$ below which phase-space density is discarded to save on memory usage and computational load \citep{Palmroth2018}. 

The simulation is driven by solar wind incoming from the $+X$ wall with homogeneous and steady parameters: density of 1\,cm$^{-3}$, velocity along the $-X_\mathrm{GSE}$ direction at 750\,km~s$^{-1}$, temperature of 0.5\,MK, and purely southward interplanetary magnetic field of 5\,nT magnitude. Boundary conditions at the $-X$, $\pm Y$ and $\pm Z$ walls of the simulation domain are homogeneous Neumann conditions (i.e. gradients are specified to be equal to zero). This run has a total duration of 1506\,s, of which we will focus on the part starting at $t=900$~s since before this time the magnetotail is not yet well formed. 

\begin{figure}
  \centering
  \includegraphics[width=0.95\columnwidth]{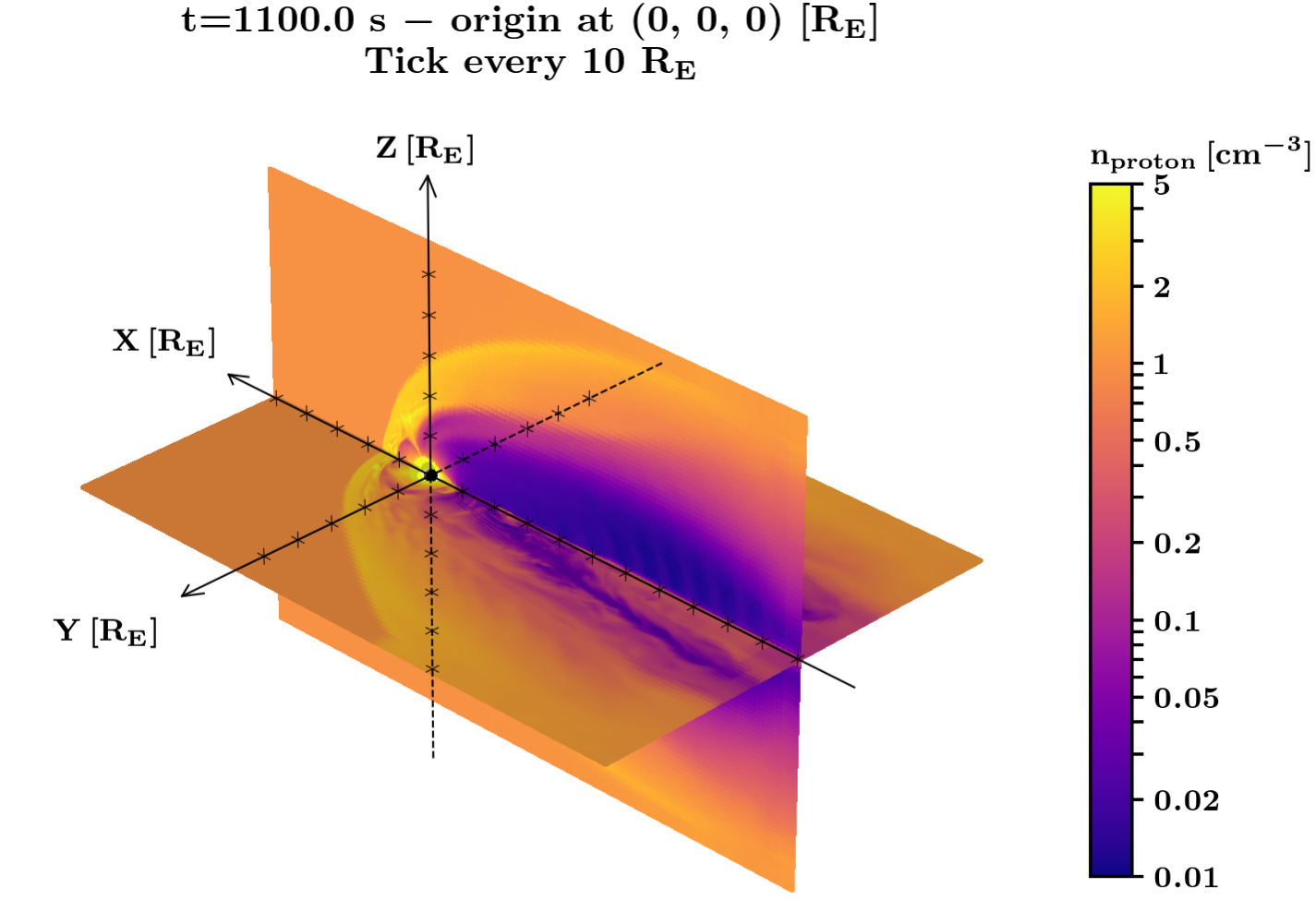}
    \caption{Proton number density in the $Y=0$ and $Z=0$ planes, covering the entire simulation domain, at $t = 1100$~s. Positive $Y$ values are towards dusk.}
    \label{fig:fig1}
\end{figure}

Figure~\ref{fig:fig1} shows an overview of the plasma density in the entire simulation domain consisting of two slices (in the $Y=0$ and $Z=0$ planes) at $t=1100$\,s in this simulation. This time corresponds to the start of active reconnection in the near-midnight magnetotail. One can identify the large-scale features of near-Earth space (bow shock, magnetopause, northern polar cusp, magnetotail) and notice the presence of structures in the dayside magnetosheath -- associated with mirror-mode waves -- as well as in the tail current sheet.

\begin{figure}
  \centering
  \includegraphics[width=0.95\columnwidth]{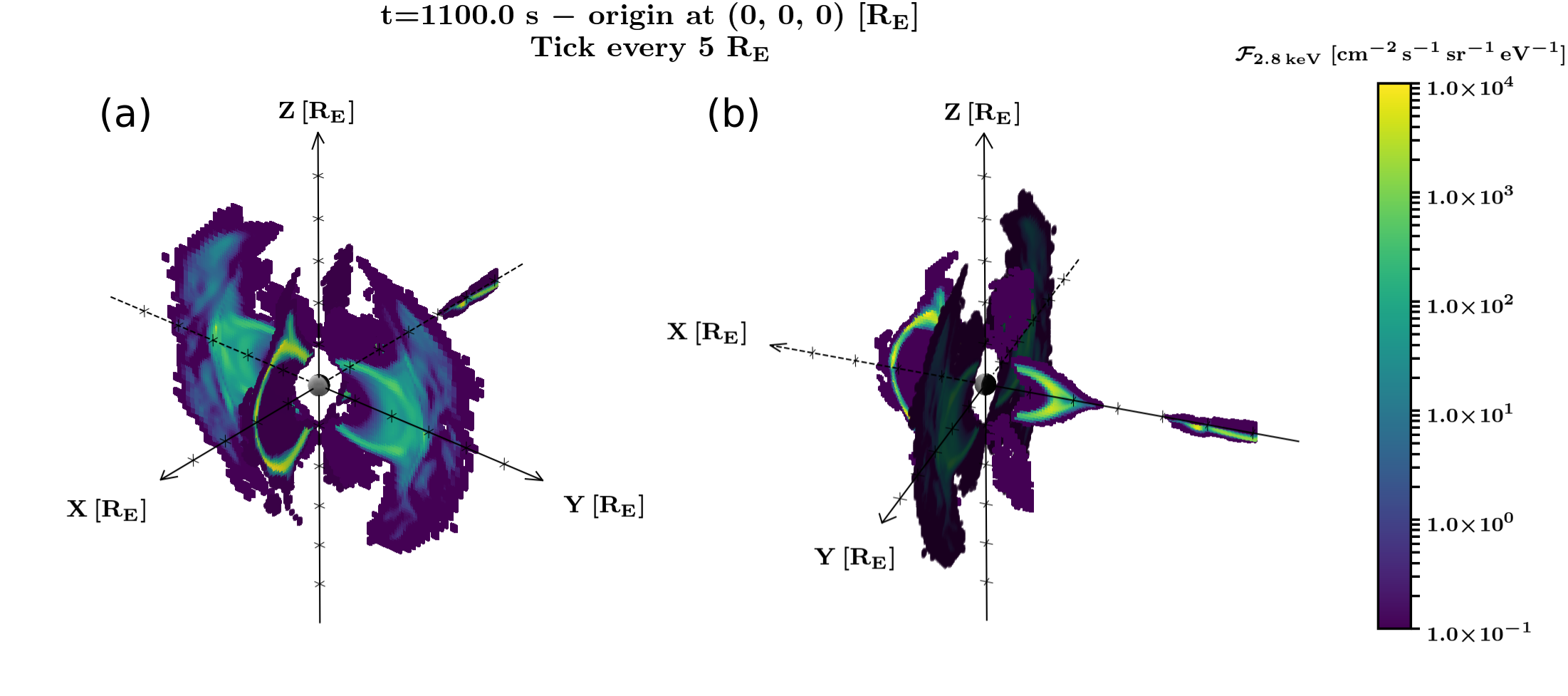}
    \caption{Directional differential number flux of protons at 2.8\,keV in the $X=0$ and $Y=0$ planes at $t=1100$\,s. The evaluation direction is within 10 degrees parallel to the local magnetic field in the $Z>0$ half of the domain and antiparallel to the local magnetic field in the $Z<0$ half. (a)~View on the dayside. (b)~View on the nightside. Positive $Y$ values are towards dusk.}
    \label{fig:fig2}
\end{figure}

In this 3D--3V run, the method presented in \citet{Grandin2019} and used to calculate the precipitating proton differential number flux from the proton velocity distribution was implemented into the Vlasiator code itself. Previous precipitation studies using Vlasiator required the differential fluxes to be evaluated a posteriori from output files, which was possible only at a few locations where the full VDF is saved in the output files and therefore limited the applicability of the method. With the implementation of the calculation directly into the Vlasiator code, precipitating fluxes can be obtained from all the cells within the simulation domain. In practice, when Vlasiator was being run to produce this simulation, the differential number flux of protons was evaluated in each cell within a cone of $10^\circ$ angle centred around the parallel (antiparallel) local magnetic field direction in the northern (southern) half of the simulation domain, at each time step and in 9~energy bins logarithmically spaced from 0.5 to 50\,keV. Note that this directional differential number flux can be interpreted as a precipitating flux only at locations (i) which are magnetically connected to the Earth and (ii) where the true loss cone angle value is not too different from $10^\circ$ (see Sect.~\ref{sec:limitations}). Figure~\ref{fig:fig2} shows the directional differential flux thus obtained for 2.8\,keV protons in the $X=0$ and $Y=0$ planes of the simulation domain at $t=1100$\,s. In Figure~\ref{fig:fig2}a, one can identify precipitating particles propagating along the dayside magnetopause and into the polar cusps, as well as less intense precipitation signatures in the dawn and dusk sectors. The second viewing angle shown in Figure~\ref{fig:fig2}b enables the examination of precipitation on the nightside. It reveals that precipitating protons originating from the magnetotail reconnection line in the midnight sector are travelling towards both the northern and the southern hemispheres. In both panels, blank locations correspond to cells where the bounce loss cone is empty and the precipitating flux is therefore zero.

\subsection{Observational data}
Observational data come from measurements of precipitating protons fluxes by the Defense Meteorological Satellite Program (DMSP) Special Sensor J (SSJ) instrument \citep{Redmon2017}. DMSP satellites fly on polar, Sun-synchronous low-Earth orbits and provide particle fluxes along their track, enabling the observation of precipitating particles in the polar region of each hemisphere at fixed local times. The spacecraft are orbiting at about 840~km altitude and have an orbital period on the order of 101~min.

The current version of the SSJ instrument is SSJ/5, which provides particle counts within a field of view spanning \mbox{$4^\circ \times 90^\circ$} in the observation plane \citep{Hardy2008}. Differential number fluxes of protons are obtained in 19~logarithmically spaced energy bins within the 30\,eV--30\,keV range. The integrated energy flux and mean precipitating proton energy are calculated from the differential number flux.

The events used for comparison with Vlasiator proton precipitation fluxes took place on 1~August 2011 and 10~October 2015, for which SSJ observations from three DMSP satellites are available: F16, F17, and F18. The SSJ data were retrieved from the CEDAR Madrigal database (\url{http://cedar.openmadrigal.org/}).

Finally, in order to find those two events for comparison, solar wind parameters and geomagnetic indices were obtained from the OMNI database at 5-min resolution \citep{King2005,OMNI_data_5min}.

\section{Method for Vlasiator precipitation mapping}
\label{sec:mapping}

Given that the inner boundary of the Vlasiator simulation lies at 4.7~$R_\mathrm{E}$, the precipitating proton fluxes cannot be evaluated immediately above the ionosphere, where low-Earth orbiting spacecraft typically measure them. Rather, we extract the fluxes in the vicinity of the inner boundary and map them to ionospheric altitudes along the magnetic field, assuming that Liouville's theorem holds, i.e. that no field-aligned potential drops significantly affect them during the propagation. While field-aligned potential drops are crucial in the formation of discrete electron aurora, it is reasonable to neglect their effect on precipitating protons of keV energies \citep[e.g.][]{Liang2013}.

Since the geomagnetic field between the Earth and the inner boundary in the Vlasiator simulation consists only of the unperturbed dipole field, we use the Tsyganenko 2001 (T01) model \citep{Tsyganenko2002a,Tsyganenko2002b} to account for the external contributions to the field in this domain. In practice, this is done using the Python Geopack library developed by Sheng Tian \citep{geopack}. First, a grid in geomagnetic latitude (MLAT) and magnetic local time (MLT) is built at auroral altitudes in the ionosphere (we use an altitude of 110~km, where DMSP/SSJ observations are mapped to), in both hemispheres. From each grid point, we follow the geomagnetic field line obtained by combining the untilted dipole model for internal contributions with the T01 model for external contributions.

Most of the parameters needed for the T01 model are directly obtained from the Vlasiator driving conditions: dynamic pressure of 1.1\,nPa, zero IMF $B_y$, IMF $B_z$ of $-5$\,nT. The indices called G1 and G2 used in the tail field parametrisation \citep[which depend on the solar wind speed and on the IMF;][]{Tsyganenko2002b} get values of 10.4 and 18.8, respectively. We use a Dst index value of $-30$\,nT, based on the assessment that the geomagnetic conditions during this Vlasiator run correspond to a relatively weak storm (this assumption is further discussed in Sect.~\ref{sec:DMSPcomparison}).

\begin{figure}
  \centering
  \includegraphics[width=0.85\columnwidth]{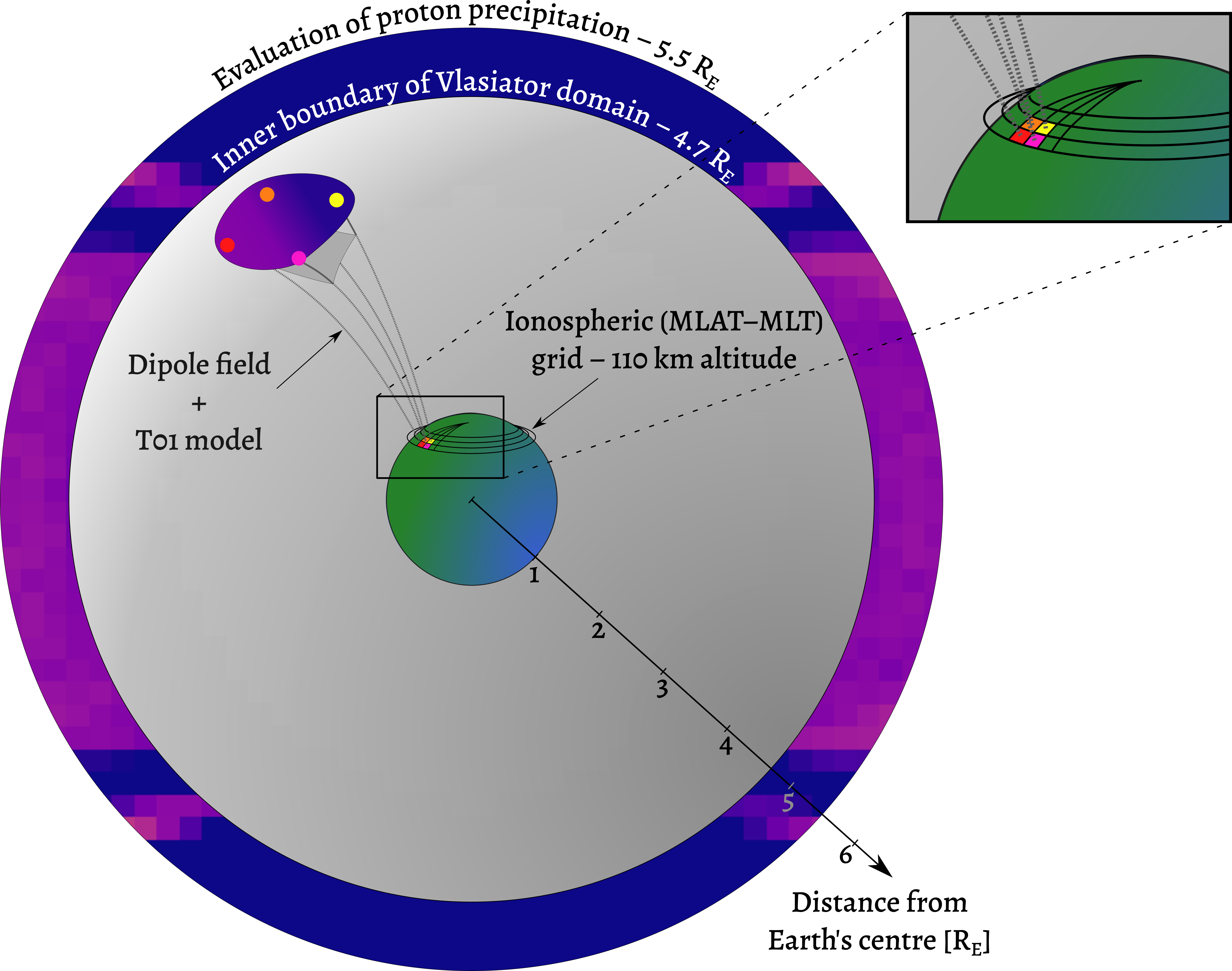}
    \caption{Illustration of the mapping of the proton precipitation between ionospheric altitudes and the Vlasiator simulation domain. The coloured cells in the MLAT--MLT ionospheric grid match the dots inside the Vlasiator domain at $5.5\,R_\mathrm{E}$ distance from the Earth's centre, where the precipitating fluxes are evaluated. The dashed (solid) grey lines joining them represent geomagnetic field lines consisting of an ideal geodipole field with external contributions given by the T01 model inside (outside of) the inner boundary of the Vlasiator domain at $r = 4.7\,R_\mathrm{E}$.}
    \label{fig:fig3}
\end{figure}

Geomagnetic field lines thus defined, with seed points on the ionosphere grid, are followed until a distance of 5.5~$R_\mathrm{E}$ from the Earth centre, where precipitating proton fluxes are available from Vlasiator outputs far enough from the inner boundary to avoid potential boundary effects. The differential fluxes of precipitating protons are then taken from the spherical shell at 5.5~$R_\mathrm{E}$ and mapped to the auroral zone and polar cap at 110~km altitude (see Fig.~\ref{fig:fig3} for an illustration of the mapping). A direct consequence of having the fluxes taken at 5.5~$R_\mathrm{E}$ from the Earth centre is that the mapped precipitation region extends only to latitudes polewards from $\pm63^\circ$, which does therefore not encompass the full auroral zone, especially when geomagnetic conditions are enhanced. This limitation can however be circumvented in future runs, in which the Vlasiator inner boundary will be pushed Earthwards as computational constraints are overcome.

Finally, we account for the time delay between the moment when precipitating protons are evaluated at 5.5~$R_\mathrm{E}$ and when they are expected to reach ionospheric altitudes by estimating their time of flight (which is different for each proton energy bin) from the length of the field line obtained during the mapping procedure. This time of flight is on the order of 10--100\,s, depending on the energy of the precipitating protons in the 0.5--50~keV energy range.

\section{Results}
\label{sec:results}

\subsection{Global proton precipitation patterns}

\begin{figure}
  \centering
  \includegraphics[width=0.95\columnwidth]{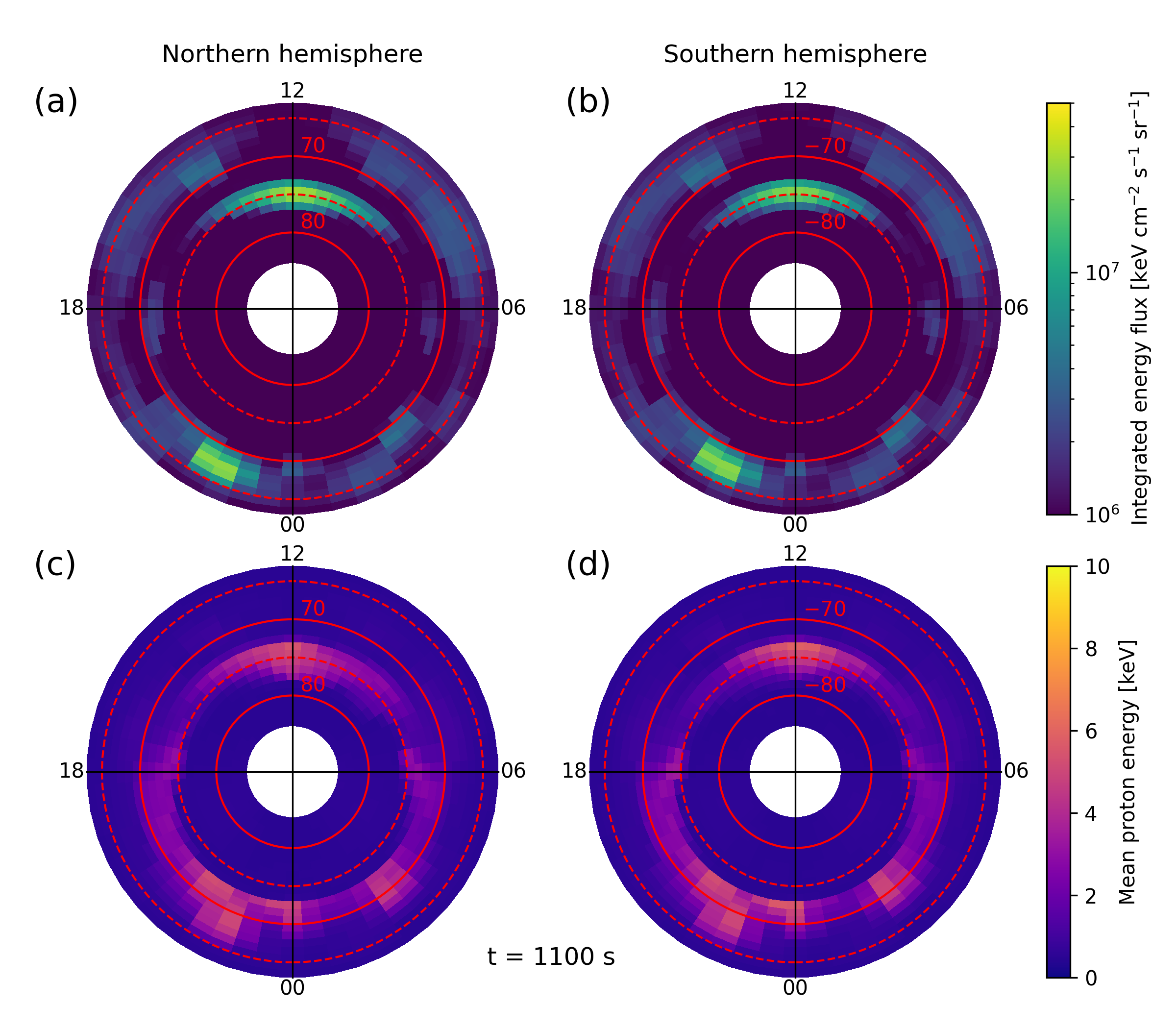}
    \caption{Overview of proton precipitation in the polar regions at $t=1100$\,s in the simulation. (a)~Precipitating proton integrated energy flux in the northern hemisphere. (b)~Same in the southern hemisphere. (c)~Precipitating proton mean energy in the northern hemisphere. (d)~Same in the southern hemisphere. In each panel, the radial coordinate is geomagnetic latitude, and the angular coordinate corresponds to MLT (noon on top). The red circles indicate selected geomagnetic latitudes.}
    \label{fig:dial2x2}
\end{figure}

We first examine the proton precipitation in the Vlasiator simulation in terms of global patterns. Figure~\ref{fig:dial2x2} shows the integrated energy flux of precipitating protons (panels a and b) and the mean precipitating proton energy (panels c and d) at $t = 1100$~s in the simulation in the polar region of the northern (panels a and c) and southern (panels b and d) hemispheres. In each panel, the geomagnetic pole is located at the centre, and geomagnetic latitudes are indicated with concentric red circles, down to $\mathrm{MLAT} = \pm65^\circ$. The angular coordinate corresponds to the magnetic local time, with noon at the top and midnight at the bottom; both hemispheres are displayed such that dawn is on the right and dusk on the left to ease the comparison between panels.

In both hemispheres, the maximum integrated energy flux values are reached in the cusp, near 12~MLT and around $\mathrm{MLAT} = \pm 75^\circ$, and on the nightside near 22~MLT and at $\mathrm{MLAT} = \pm 67^\circ$. The location of these two maxima exhibits nearly perfect symmetry between both hemispheres, especially in the nightside. The nightside spot peaks at a value of $2.5 \times 10^7$~keV\,cm$^{-2}$\,s$^{-1}$\,sr$^{-1}$, in both hemispheres,
and it has a quite reduced extent in MLT. The fact that the nightside spot is localised in MLT can be explained by taking into account the magnetic reconnection process occurring in the magnetotail.

Supplementary Animation~SA1 presents an animated version of Figure~\ref{fig:dial2x2}, with an additional panel at the bottom showing the Sunward component of the plasma bulk velocity, $V_x$, in the $Z=0$ plane in the near-Earth magnetotail. Animation~SA1 visualises these data during the entire studied time interval, i.e. from $t = 900$ until $t = 1506$~s. Magnetotail magnetic reconnection is continuously ongoing during the simulation, as suggested by the uninterrupted presence of a region of flow reversal ($V_x=0$) at $X \sim -13\,R_\mathrm{E}$, extended along the $Y$ direction and corresponding to the reconnection X-line. As expected during magnetotail reconnection, plasma is diverted and accelerated in the outflow regions. The tailward outflow region is characterised by $V_x < 0$ reaching values up to $V_x \sim -1000$~km/s, and the Earthward outflow region presents $V_x > 0$ reaching maximum values comparable to the tailward outflow. The magnetotail is highly dynamic and the reconnection X-line has a complex structure. As reconnection is not steady, it produces bursty bulk flows \citep[BBFs; e.g.][]{Angelopoulos1994,Baumjohann1990}. A BBF can for instance be identified in the bottom panel of Supplementary Animation~SA1 between $t \approx 970$ and $t \approx 1090$ s as the region of enhanced Earthward flow located at $X > -10\,R_E$ and $Y = 5\,R_E$. The nightside precipitation spot visible in Figure~\ref{fig:dial2x2} is likely associated with this BBF, given the time needed for precipitating protons originating from the tail region to reach 110~km altitude in the ionosphere.

The main difference between the northern and the southern hemispheres can be found by examining closely the cusp integrated energy flux: while the maximum flux value in the northern hemisphere is $3.0 \times 10^7$~~keV\,cm$^{-2}$\,s$^{-1}$\,sr$^{-1}$, it is lower in the southern hemisphere ($2.4 \times 10^7$~~keV\,cm$^{-2}$\,s$^{-1}$\,sr$^{-1}$). This is consistent with earlier findings in 2D, where it was shown that cusp proton precipitation takes place in the form of successive bursts corresponding to the transit of flux transfer events (FTEs) in the high-altitude cusp, under southward IMF driving, hence not occurring in a symmetrical way in both hemispheres \citep{Grandin2020}. The relation between these cusp proton precipitation bursts and FTEs will be investigated in Sect.~\ref{sec:spectra}.

Looking at the mean proton energy (Figs~\ref{fig:dial2x2}c and d), we find that regions characterised with enhanced integrated energy flux are associated with mean precipitating proton energies ranging within 4--6\,keV. Moreover, in the nightside, regions of the auroral oval away from the main precipitation spot, where the integrated energy flux remains under $1 \times 10^7$~keV\,cm$^{-2}$\,s$^{-1}$\,sr$^{-1}$, have mean proton energies on the same order. This suggests that keV protons are scattered in the magnetotail although reconnection is not as active as near the aforementioned fast Earthward flow at corresponding tail locations (see Animation~SA1). Proton precipitation in this case is probably due to the loss-cone scattering of those keV protons through the field line curvature process \citep{Tsyganenko1982,Sergeev1982}. This assumption is supported by Fig.~SF1 in the supplementary material, assessing in the near-Earth current sheet the criterion defined by \citet{Sergeev1983} for loss-cone scattering to occur. This criterion is formulated as $\kappa \leq \sqrt{8}$, where $\kappa = \sqrt{R_c / r_L}$, with $R_c$ the magnetic field curvature radius and $r_L$ the gyroradius of the particle. The criterion is verified across the nightside MLTs in a band located at $X \approx -11\,R_\mathrm{E}$ within $|Y| \leq 5\,R_\mathrm{E}$. 

Again, the nightside precipitation exhibits very strong symmetry between both hemispheres, while slight differences can be seen in the cusp precipitation -- this time, the southern hemisphere cusp having slightly larger mean energy values than the northern hemisphere cusp. The animation shows that nightside precipitation patterns remain symmetric throughout the time interval, but exhibit noticeable temporal variations. Shortly after $t=1100$~s, proton precipitation takes place in the midnight MLT sector. It is likely associated with the enhancement in the Earthward flow which penetrates closer to the Earth (around $Y=0$ and at $X>-10\,R_E$) starting from $t \approx 1030$~s in the simulation. For the rest of the simulation, precipitating protons are present in the 22--1~MLT sector, and active reconnection is ongoing in the near-Earth tail. On the dayside, the cusps exhibit proton precipitation bursts in both hemispheres. The bursts are not correlated between the hemispheres. It will be shown in the next section that this is a consequence of cusp precipitation being associated with the transit of FTEs in the cusps, which occurs asynchronously in both hemispheres.

\subsection{Local precipitating proton energy spectra}
\label{sec:spectra}

In order to investigate the characteristics of the auroral proton precipitation in more detail, we monitor the energy spectrum of precipitating protons at selected locations, first in the cusp and then in the nightside oval. Given that the northern and southern hemispheres exhibit similar behaviours, we will only show results from the northern hemisphere in this subsection.

\begin{figure}
  \centering
  \includegraphics[width=0.95\columnwidth]{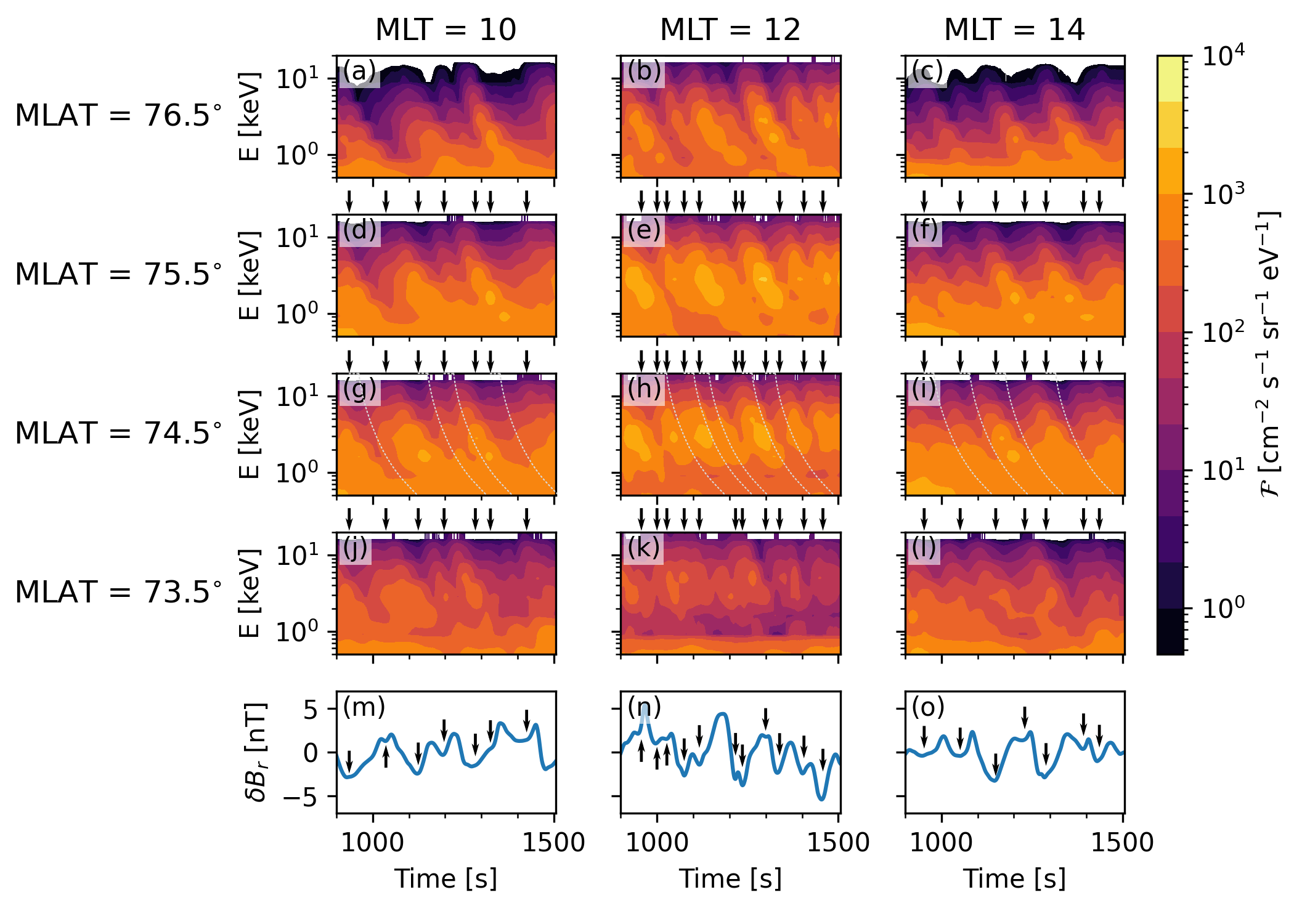}
    \caption{(a--l)~Precipitating proton energy spectrum (i.e. differential number flux) at selected locations in the northern-hemisphere cusp region as a function of time in the simulation. (m--o) Temporal variations in the radial magnetic field component at three virtual spacecraft placed in the high-altitude cusp at 10, 12 and 14~MLT. Black arrows indicate times preceding signatures of the leading edge of incoming FTEs and are reproduced above panels d--l. The thin dashed lines in panels g--i indicate the theoretical time of detection of enhanced precipitating proton flux, as a function of proton energy, following the arrival of selected FTEs in the high-altitude cusp.}
    \label{fig:spectra_cusp}
\end{figure}

Figure~\ref{fig:spectra_cusp} shows the time evolution of the differential number flux of precipitating protons around the cusp, at three MLTs (10, 12, and 14, organised in columns) and four MLATs (76.5$^\circ$, 75.5$^\circ$, 74.5$^\circ$, and 73.5$^\circ$, organised in rows). In Figure~\ref{fig:spectra_cusp}a--l, signatures of individual precipitation bursts can be identified. They all exhibit an energy dispersion, with high energies arriving first and low energies arriving last (this is particularly visible in panels a--h). At a given MLT, each same burst is generally detected at 74.5$^\circ$, 75.5$^\circ$ and 76.5$^\circ$ MLAT, in this order. These signatures are less clear at 73.5$^\circ$ MLAT (Figs~\ref{fig:spectra_cusp}j--l), albeit still distinguishable. We can distinguish 4~bursts at 10~MLT, 7 at 12~MLT, and 5 at 14~MLT, and some signatures seem to coincide in time at two, sometimes three (e.g. the one starting at high energies shortly after $t = 1100$~s) MLTs. Therefore, the MLT extent of the cusp (in terms of proton precipitation signature) varies throughout the simulation, as can also be seen in Supplementary Animation~SA1. The largest differential number flux values are reached at 12~MLT and 74.5--75.5$^\circ$~MLAT, which indicates that cusp fluxes peak around magnetic noon. This is consistent with the driving conditions with purely southward IMF (no $B_y$ component). The count of 7~auroral proton precipitation bursts taking place over 606~s (Fig.~\ref{fig:spectra_cusp}e) yields a burst occurrence rate of about 0.7~burst per minute (i.e. one burst every 87~s on average).

To investigate a possible association of precipitation bursts with the transit of FTEs in the cusp, we examine the temporal variations of the projection of the radial component of the magnetic field in the XZ plane (with respect to the centre of the Earth; see Supplementary Figure~SF2 for a sketch), denoted $B_r$, at three fixed locations that will act as virtual spacecraft. These virtual spacecraft are placed in the high-altitude cusp on field lines having their ionospheric footpoints at 10, 12 and 14~MLT. They are located at $(X_\mathrm{GSE}, Y_\mathrm{GSE}, Z_\mathrm{GSE}) = (4.67, -2.70, 7.50)\,R_\mathrm{E}$, $(5.50, 0.00, 7.50)\,R_\mathrm{E}$, and $(4.67, 2.70, 7.50)\,R_\mathrm{E}$, respectively. After subtracting the time-averaged value within $t=900$--1506~s, $\bar{B_r}$, we show $\delta B_r = B_r - \bar{B_r}$ in Figure~\ref{fig:spectra_cusp}m--o. When a FTE reaches the northern polar cusp, its leading edge produces an increase in the radial magnetic field component (see Supplementary Figure~SF2). We indicate with black arrows the times when such an increase in $\delta B_r$, which we interpret as the signature of an incoming FTE in the high-altitude cusp, is detected at the virtual spacecraft locations in Figure~\ref{fig:spectra_cusp}m--o, and we reproduce such arrows above panels d--l. To aid visual inspection, we add a few light-grey dashed lines to indicate around what time we can expect an enhancement in the precipitating fluxes following the arrival of selected FTEs, in panels~g--i. This arrival time corresponds to an energy-dependent delay added to the time of the corresponding arrow, as the FTE detection takes place in the high-altitude cusp, whereas precipitating fluxes are time-propagated to 110~km altitude. For 20~keV protons, the delay is on the order of 30~s, whereas for 0.5~keV protons it is 195~s. Although there are slight discrepancies -- likely due to the uncertainty regarding the exact location at which the FTE is appended to lobe field lines polewards of the cusp -- we can clearly see that, at all three MLTs, the times indicated with black arrows are generally followed by an enhancement in precipitating proton flux exhibiting the expected temporal energy dispersion. The same analysis is carried out for the southern hemisphere and shown in Supplementary Figure~SF3; it leads to the same observations. This therefore confirms that cusp precipitation bursts are associated with FTEs, and an additional illustration of this process is provided as Supplementary Figure~SF4 showing the appending of a FTE to lobe field lines in the $Y=0$ plane (12~MLT on the dayside) at three time steps. While a more detailed study of how FTEs interact with lobe field lines is beyond the scope of this paper, we note that it was found from a 2D--3V Vlasiator simulation that FTEs provide a significant contribution to the energy input into the magnetosphere by reconnecting with lobe field lines polewards of the cusp \citep{AlaLahti2022}.

\begin{figure}
  \centering
  \includegraphics[width=0.95\columnwidth]{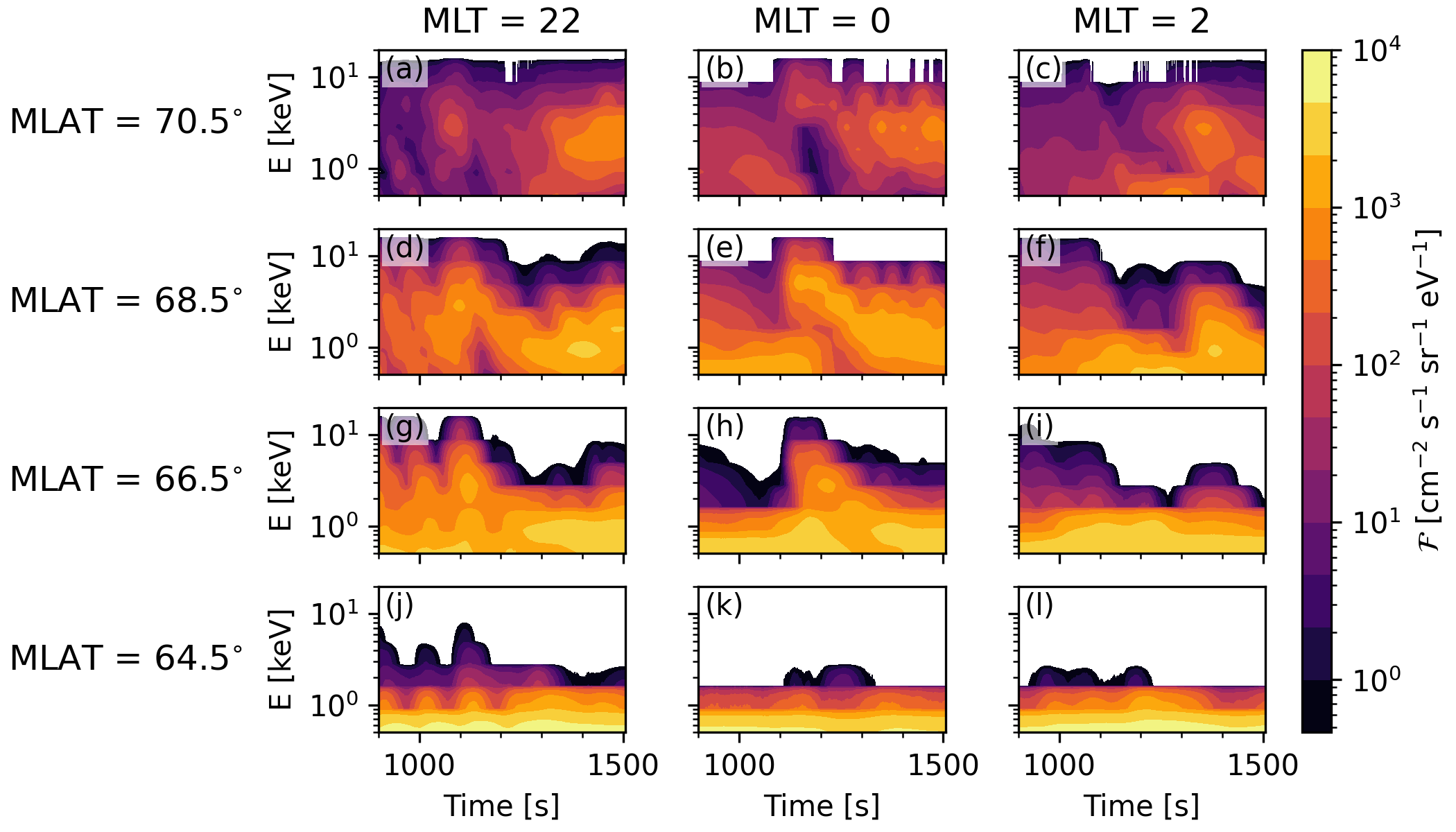}
    \caption{Precipitating proton energy spectrum at selected locations in the nightside northern-hemisphere auroral oval region as a function of time in the simulation.}
    \label{fig:spectra_night}
\end{figure}

Figure~\ref{fig:spectra_night} shows the time evolution of the differential number flux of proton precipitation in the nightside oval (northern hemisphere), in the same format as the top panels of the previous figure. Here, we consider MLTs of 22, 0 and 2, and the selected MLATs are 70.5$^\circ$, 68.5$^\circ$, 66.5$^\circ$, and 64.5$^\circ$. The precipitation signatures show a different behaviour compared to the dayside, with steadier spectra not necessarily exhibiting a clear energy dispersion (see for instance between $t = 1300$ and $t = 1500$~s in Figs~\ref{fig:spectra_night}e and g). Spectra are in addition relatively different across MLTs, with no clear simultaneous signatures at two different MLTs. This suggests that precipitating proton sources in the tail have a limited extent along the $Y$ direction, which was already noted when discussing Supplementary Animation~SA1 in the previous section. However, in the same MLT sector, individual signatures can generally be observed over a range of MLATs (see for instance the three bursts of precipitation prior to $t = 1200$~s at 22~MLT, or the feature starting around $t = 1120$~s at 0~MLT). Overall, very little proton precipitation above 2~keV is seen at the most equatorward MLAT (Figs~\ref{fig:spectra_night}j--l), which suggests that the equatorward edge of the auroral proton oval lies above 64.5$^\circ$~MLAT. We also note that there is no significant flux of precipitating protons at energies greater than 16~keV in the entire nightside oval (the next two energy bins being 28 and 50~keV, due to the logarithmic distribution of the 9~central energies for the evaluation of the differential number flux in this run).

\subsection{MLAT variations of proton precipitation at fixed MLTs}

In this subsection, we focus on the MLAT dependency of proton precipitation features, again both in the dayside and in the nightside, at fixed MLTs. To that end, we examine differential number fluxes at two selected energies encompassing the main range of auroral proton energies in this run: 1.6~keV and 8.9~keV.

\begin{figure}
  \centering
  \includegraphics[width=0.95\columnwidth]{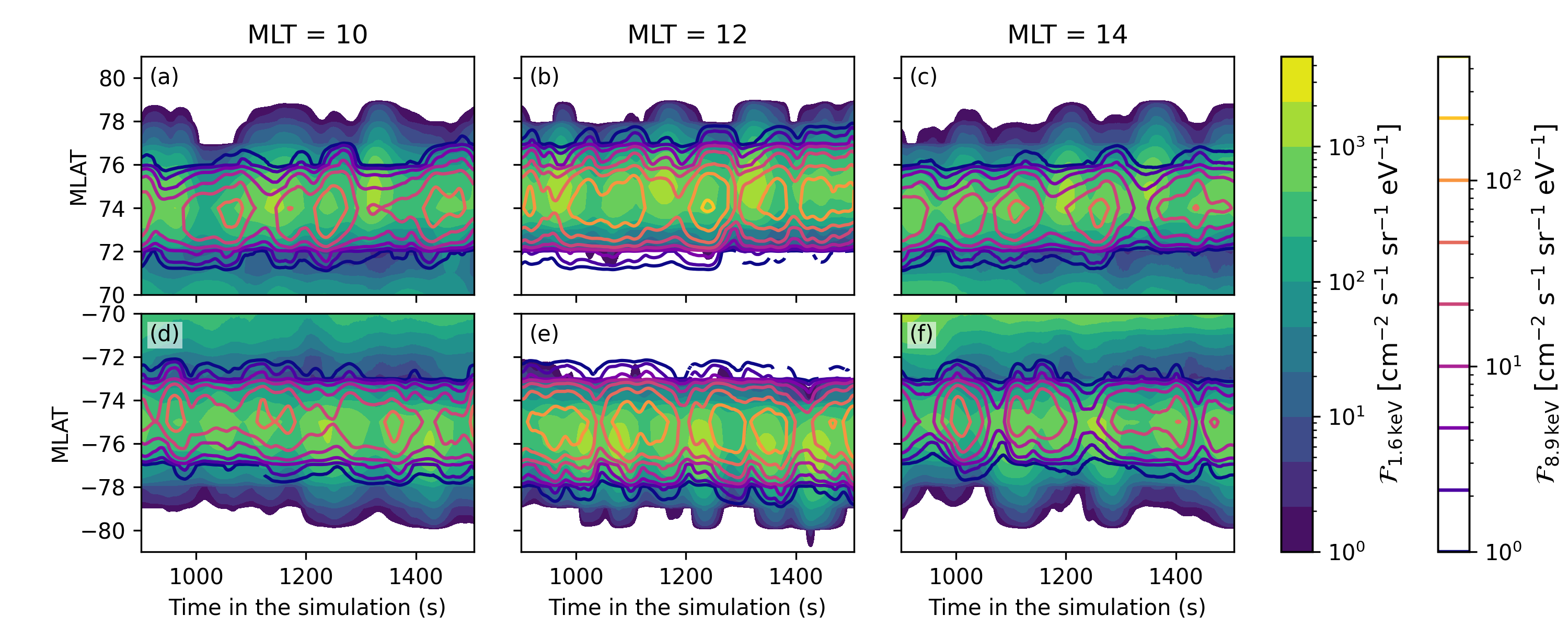}
    \caption{Differential number flux of precipitating protons at 1.6\,keV (background colour) and at 8.9\,keV (isocontour lines) as a function of time in the simulation and geomagnetic latitude (MLAT) at selected magnetic local times (MLT) on the dayside. Panels a--c show the northern hemisphere, whereas panels d--f show the southern hemisphere. Note that the colour scales corresponding to the two energy bins differ from each other.}
    \label{fig:mlatkeo_day}
\end{figure}

Figure~\ref{fig:mlatkeo_day} shows the time evolution of the differential number flux of precipitating protons at those two energies as a function of MLAT, at three MLTs on the dayside: 10, 12, and 14 (like in Fig.~\ref{fig:spectra_cusp}). The 1.6~keV flux is shown with the background colour, whereas the 8.9~keV flux is shown with the isocontours. Panels a--c correspond to the northern hemisphere, and panels d--f to the southern hemisphere. In this figure too, the burst-like signatures associated with cusp precipitation are well visible, in all panels. The 8.9~keV flux peaks at MLATs slightly equatorwards from the peak location of the 1.6~keV flux, which indicates a latitudinal energy dispersion consistent with earlier findings in case of southward IMF driving \citep{Tan2012,Connor2015,Grandin2020}. Besides this spatial energy dispersion, the temporal energy dispersion discussed in Sect.~\ref{sec:spectra} is also visible, as the burst signatures at 8.9~keV and at 1.6~keV do not occur at the same time (visible in all panels). In addition, the asynchronous nature of cusp precipitation in the northern and southern hemisphere, already mentioned previously, is also clear from this figure, as there is no north--south symmetry when comparing panels a--c with panels d--f. Besides, it can be noted from this figure that, in a given cusp and at a given MLT, the precipitating proton fluxes are not distributed symmetrically around their peak along the north--south direction. At equal distance in MLAT from the peak, the lower latitudes tend to have lower fluxes. Finally, this figure enables the monitoring of the cusp motion in terms of MLAT as a function of time. It can be seen that, overall, the peak flux location at a given proton energy does not fluctuate much with time, suggesting that the cusp location was fairly stable throughout the simulation. This is reasonable, given that the driving conditions are steady.

\begin{figure}
  \centering
  \includegraphics[width=0.95\columnwidth]{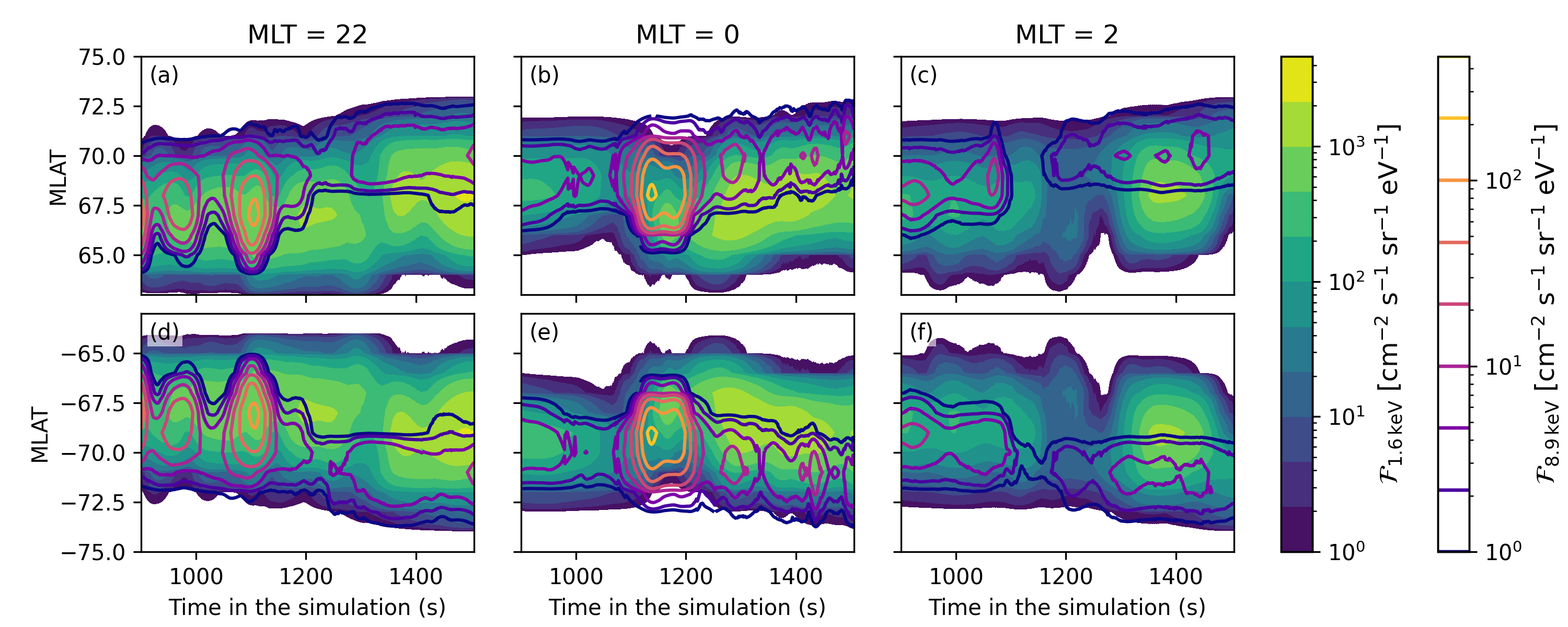}
    \caption{Differential number flux of precipitating protons at 1.6\,keV (background colour) and at 8.9\,keV (isocontour lines) as a function of time in the simulation and geomagnetic latitude (MLAT) at selected magnetic local times (MLT) on the nightside. Panels a--c show the northern hemisphere, whereas panels d--f show the southern hemisphere. }
    \label{fig:mlatkeo_night}
\end{figure}

A similar analysis is carried out on the nightside in Figure~\ref{fig:mlatkeo_night}. The format is the same as in the previous figure, except that the considered MLTs are 22, 0, and 2 (like in Fig.~\ref{fig:spectra_night}). Here, the interhemispheric symmetry is particularly visible, with differences between the northern and southern hemispheres being marginal (see for instance the 8.9~keV flux contours at 2~MLT after $t \approx 1250$~s). It is also fairly clear that the 8.9~keV precipitation is systematically found polewards from the 1.6~keV precipitation. This corresponds to what \cite{Liang2014} call a ``reversed'' energy-latitude dispersion (see Sect.~\ref{sec:discussionOtherWorks}). Yet, the most interesting feature shown by Figure~\ref{fig:mlatkeo_night} is the latitudinal motion of the precipitation in the midnight MLT sector during the simulation. Between $t \approx 1050$ and $t \approx 1150$~s, the precipitation migrates towards the equator, after which it gradually recedes polewards in MLAT and fluxes decrease, especially in the 8.9~keV energy bin. These variations are consistent with the progression of the fast Earthward flow discussed earlier, visible in Supplementary Animation~SA1, which initially moves Earthwards and then recedes slightly and dissipates. We note, further, that the nightside precipitation features differ greatly at the three shown MLTs, underlining the importance of having a 3D description of near-Earth space to capture this variability.

\subsection{MLT variations of proton precipitation at fixed MLATs}

\begin{figure}
  \centering
  \includegraphics[width=0.95\columnwidth]{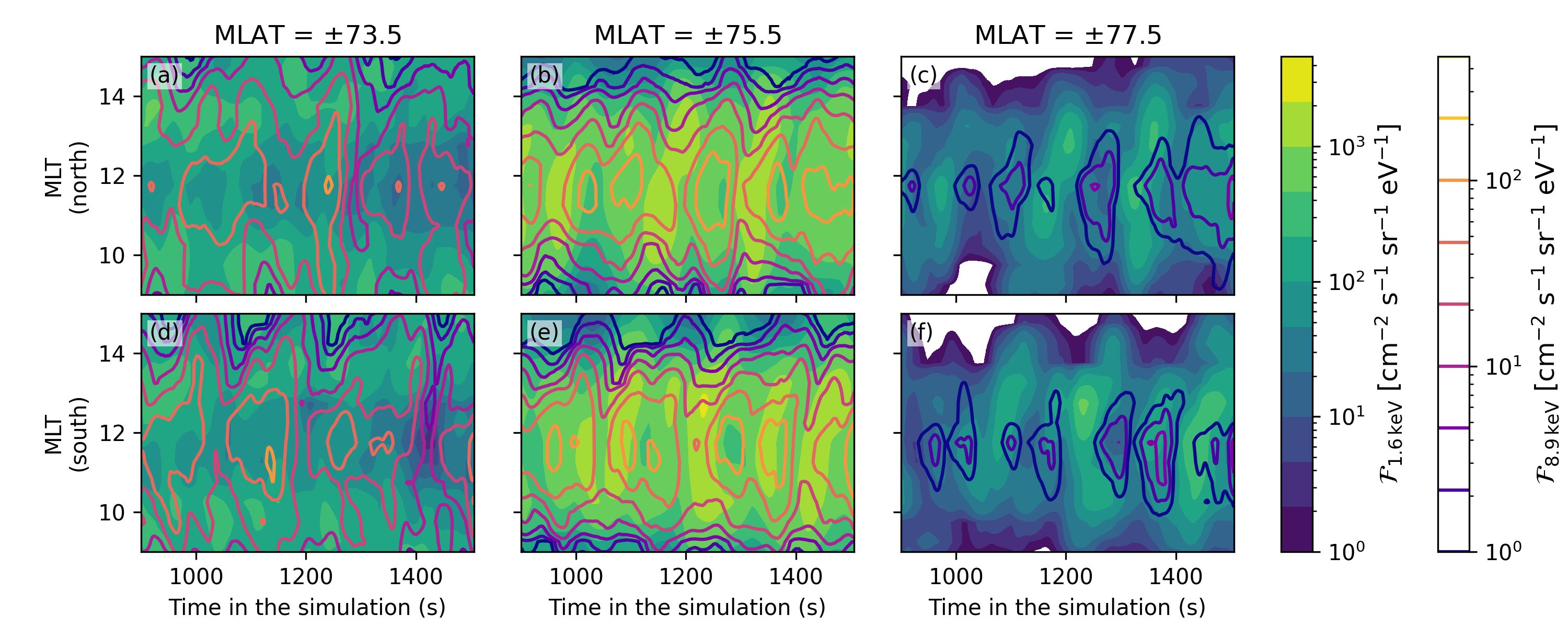}
    \caption{Differential number flux of precipitating protons at 1.6\,keV (background colour) and at 8.9\,keV (isocontour lines) as a function of time in the simulation and magnetic local time (MLT) at selected geomagnetic latitudes (MLAT) on the dayside.}
    \label{fig:mltkeo_day}
\end{figure}

We can carry out a similar analysis as above but looking at the MLT dependency of precipitating fluxes at a few selected MLATs. Figure~\ref{fig:mltkeo_day} shows the time evolution of the 1.6~keV and 8.9~keV proton precipitation on the dayside, for both hemispheres, at $\pm 73.5^\circ$, $\pm 75.5^\circ$ and $\pm 77.5^\circ$~MLAT, in an analogous format as the previous two figures, but with MLT as the vertical axis. While features already discussed in the previous subsections can again be identified (asynchronous precipitation in the northern and southern cusps, temporal energy dispersion), this figure also gives a measure of the MLT extent of the cusps, in terms of proton precipitation. Additional observations from this figure can be listed as follows:
\begin{enumerate}
\item The equatorward side of the cusp (Figs~\ref{fig:mltkeo_day}a and d) shows a different behaviour for the 1.6~keV and 8.9~keV fluxes: while the latter is mostly peaking around noon, the former actually has a local minimum around noon and increased values both in the prenoon and postnoon local times. This seems to be related to the fact that spots of $<$2~keV precipitation are present on the dayside, away from noon, throughout most of the simulation. Such spots are not present from the 2.8~keV precipitation energy bin upwards (not shown), suggesting that they might be due to a leakage of low-energy protons from the inner boundary (in the opposite hemisphere) rather than a real precipitation feature.
\item The MLT elongation of precipitating proton structures in the central cusp ($\pm 75.5^\circ$~MLAT) is larger for the 1.6~keV flux than for the 8.9~keV flux (Figs~\ref{fig:mltkeo_day}b and e): the 1.6~keV flux drops to $\sim$20\% of its peak value with a MLT change of approximately $\pm$3~h, while the 8.9~keV flux drops by a similar amount with a MLT change on the order of $\pm$2~h. 
\item The location of the peak flux values is generally close to 12~MLT, with a slight tendency to favour pre-noon MLTs. The MLT extent of such structures is relatively similar on both sides of the peak. The notable exception is the 1.6~keV flux at $\mathrm{MLAT} = \pm 73.5^\circ$ which has a local minimum around noon, as discussed above.
\end{enumerate}

\begin{figure}
  \centering
  \includegraphics[width=0.95\columnwidth]{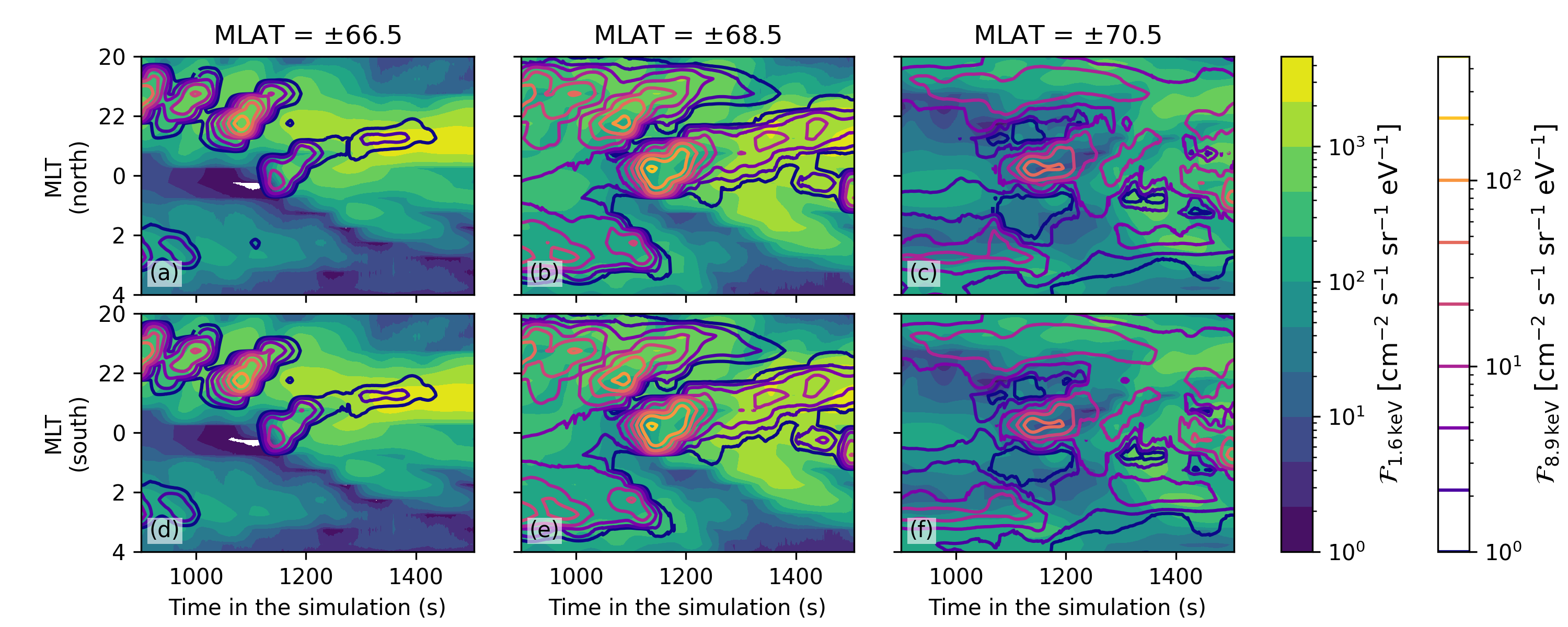}
    \caption{Differential number flux of precipitating protons at 1.6\,keV (background colour) and at 8.9\,keV (isocontour lines) as a function of time in the simulation and magnetic local time (MLT) at selected geomagnetic latitudes (MLAT) on the nightside.}
    \label{fig:mltkeo_night}
\end{figure}

The same type of plot is made for the nightside oval, this time for MLATs of $\pm 66.5^\circ$, $\pm 68.5^\circ$ and $\pm 70.5^\circ$; it is presented in Figure~\ref{fig:mltkeo_night}. The main observations are as follows:
\begin{enumerate}
\item At the lowest geomagnetic latitude (Figs~\ref{fig:mltkeo_night}a and d), three successive precipitation signatures can be identified in the pre-midnight sector between $t = 900$ and $t \approx 1125$~s, for both energies. They all start around 22~MLT and propagate towards earlier local times, which is consistent with the duskward drift of the fast Earthward flows near $Y = 5\,R_E$ in the nightside (see Supplementary Animation~SA1 from $t = 900$ to $t \approx 1090$~s). 
\item A fourth precipitation signature starts around $t = 1100$~s in the midnight sector, also exhibiting a duskward drift with time. The 1.6~keV flux reaches its maximum values from $t = 1300$~s until the end of the simulation (centred around 23~MLT), whereas at 8.9~keV the maximum values are reached at the beginning of the signature ($t \approx 1140$~s, around 0~MLT) and rapidly decrease as the precipitation drifts towards dusk. This signature is consistent with the fast Earthward flow starting near $Y=0$ around $t=1030$~s.
\item The midnight-sector precipitation signature is also present at $\pm 68.5^\circ$~MLAT (Figs~\ref{fig:mltkeo_night}b and e). The 8.9~keV (1.6~keV) flux is greater (lower) than at $\pm 66.5^\circ$~MLAT; this is consistent with the spatial energy dispersion associated with nightside proton precipitation discussed in the previous subsection.
\item The signatures are overall less clear at the highest MLAT (Figs~\ref{fig:mltkeo_night}c and f), but the near-midnight injection around $t = 1150$~s still stands out at 8.9~keV. The three pre-midnight injections prominent at $\pm 66.5^\circ$~MLAT (and, to a lesser extent, at $\pm 68.5^\circ$~MLAT) are however not present at all at $\pm 70.5^\circ$~MLAT. This is consistent with the fast Earthward flows at $Y \approx 5\,R_E$ starting closer to the Earth than the one near $Y=0$.
\end{enumerate}

\subsection{Vlasiator--DMSP/SSJ comparison}
\label{sec:DMSPcomparison}

To assess whether the proton precipitation parameters given by this Vlasiator run are in reasonable agreement with observations, we compare the integrated energy fluxes and differential number fluxes with measurements by low-Earth-orbiting satellites. We first searched for events during which the driving conditions given by the solar wind parameters (mainly density and velocity) and the interplanetary magnetic field were relatively similar to those used as an input for the Vlasiator simulation. We show here comparisons during two such events.

\begin{figure}
  \centering
  \includegraphics[width=0.95\columnwidth]{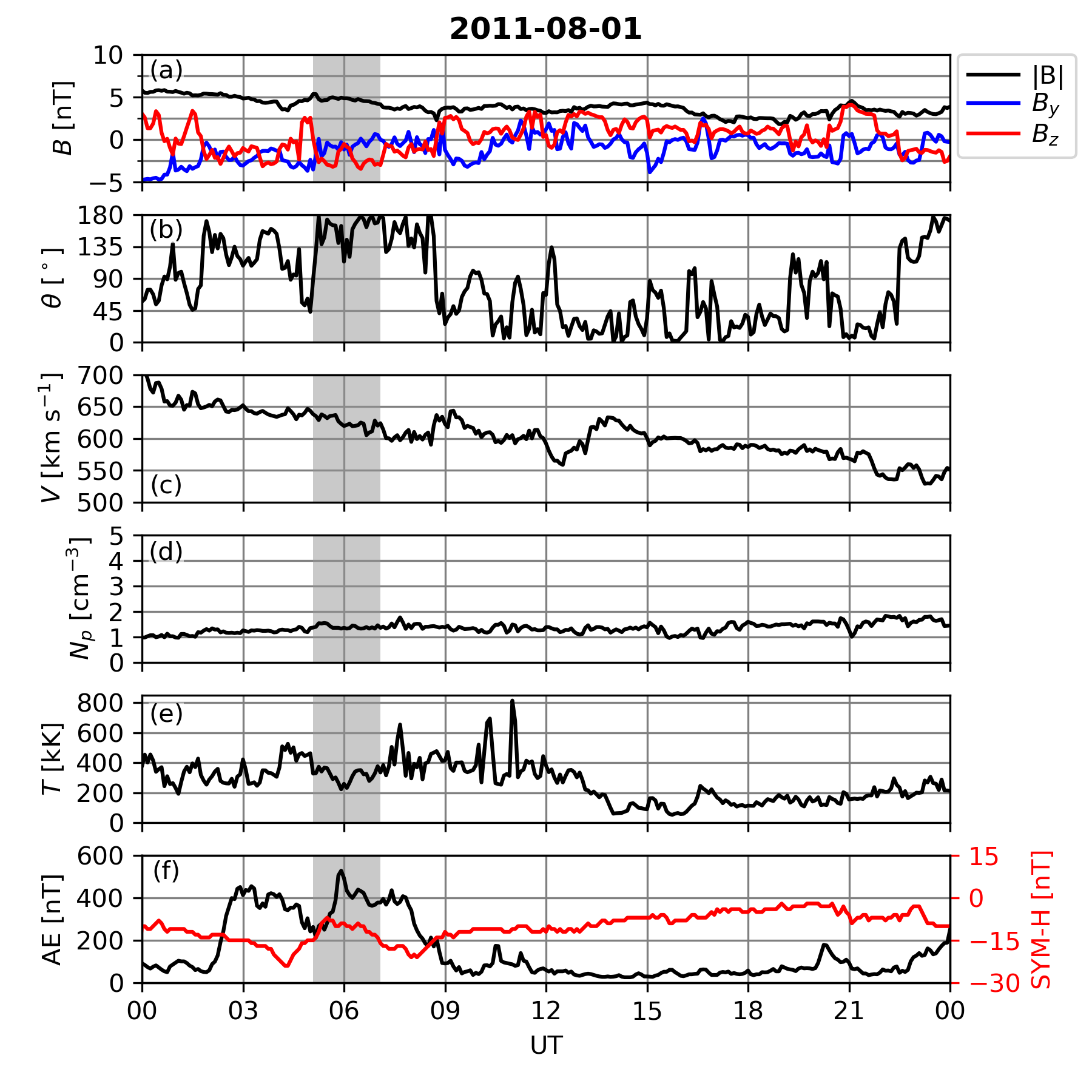}
    \caption{Solar wind conditions and geomagnetic activity on 1~August 2011, from the OMNI database. (a)~Interplanetary magnetic field (IMF), with the total field magnitude in black, and the $B_y$ and $B_z$ components in blue and red, respectively. (b)~IMF clock angle. (c)~Solar wind speed. (d)~Solar wind density. (e)~Solar wind temperature. (f)~SYM-H index (red line) and AE index (black line). The shaded area corresponds to a period during which the solar wind driving conditions were relatively similar to those used as a driver of the Vlasiator run.}
    \label{fig:2011event_omni}
\end{figure}

Figure~\ref{fig:2011event_omni} shows solar wind and geomagnetic conditions obtained from the OMNI database and measured on 1~August 2011. On that day, and especially during the early-morning hours (from 5:00 until 7:00~UT; shaded area), the IMF had a magnitude of about 5~nT, with a southward orientation during most of the time interval (clock angle close to 180$^\circ$). The solar wind speed was within 600--650~km\,s$^{-1}$, which despite being lower than in the Vlasiator run corresponds to elevated values, and the solar wind density was close to 1~cm$^{-3}$. The solar wind temperature varied between 200 and 500~kK, which is a bit less than in the Vlasiator run but still more elevated than values observed during the second half of that day.

The geomagnetic conditions associated with this solar wind driving exhibited moderately enhanced substorm activity (AE index peaking slightly over 500\,nT around 6\,UT). During the period of interest, a weak geomagnetic storm was taking place, with SYM-H values reaching $-25$\,nT around 4\,UT. This value is similar to the one used in the T01 model for the mapping of the Vlasiator inner boundary to ionospheric altitudes (see Sect.~\ref{sec:mapping}).

\begin{figure}
  \centering
  \includegraphics[width=0.95\columnwidth]{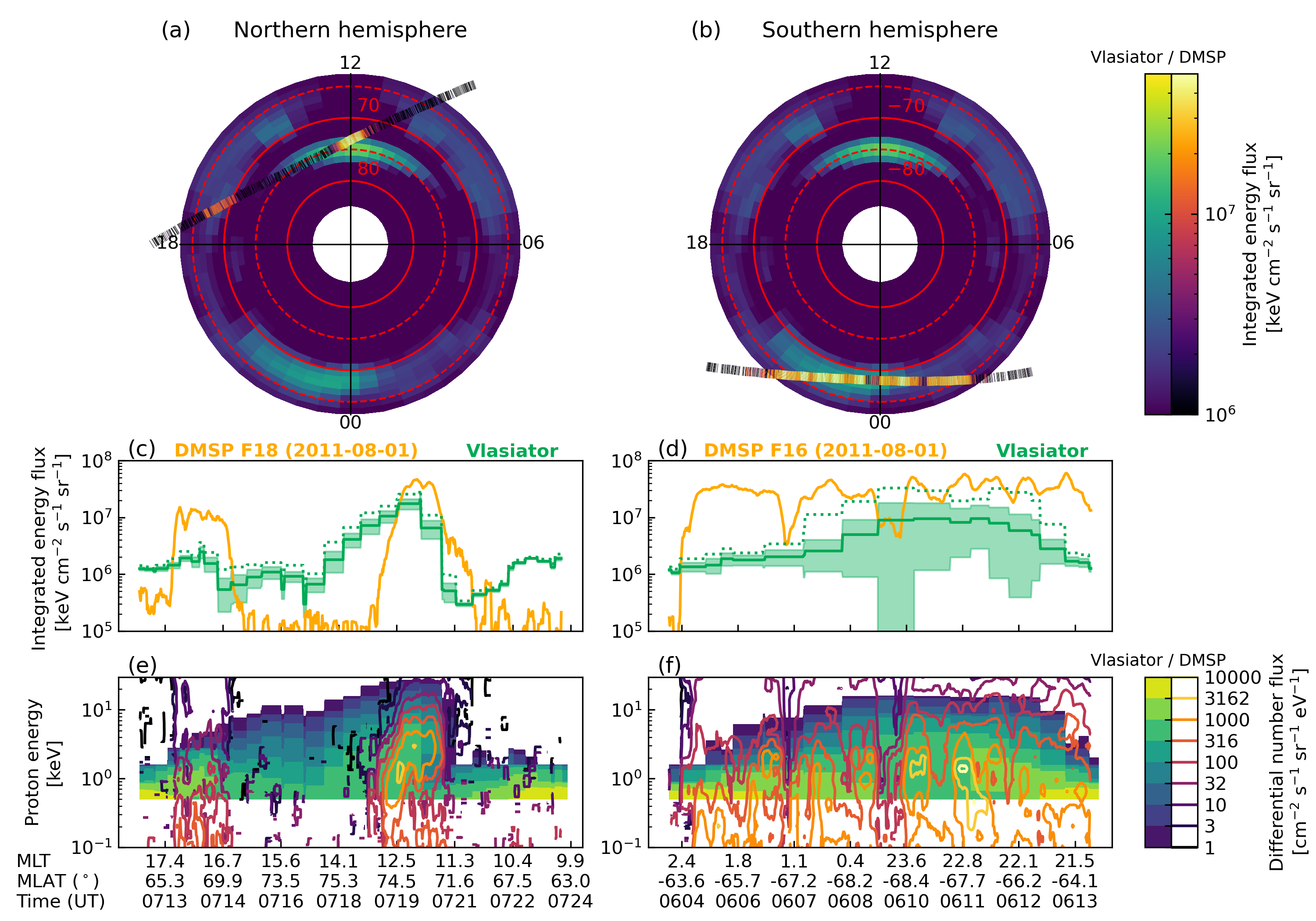}
    \caption{Comparison of Vlasiator proton precipitation and DMSP/SSJ observations on 1~August 2011. (a--b)~integrated energy flux shown in polar MLT--MLAT coordinates for Vlasiator and DMSP F18 (northern hemisphere, panel a) and DMSP F16 (southern hemisphere, panel b). The Vlasiator data correspond to time averages between $t = 900$ and $t = 1506$\,s. (c--d)~The same data along the satellite orbits, shown as time series. The orange line corresponds to the DMSP/SSJ fluxes, the solid green line gives the time-averaged Vlasiator flux, the green shaded area indicates the $\pm 1\sigma$ (standard deviation) interval, and the dotted green line shows the maximum flux value during the considered simulation time range. (e--f)~Differential number flux of precipitating protons along the satellite orbits in Vlasiator (background colours, time-averaged) and DMSP/SSJ observations (isocontours).}
    \label{fig:2011event_data}
\end{figure}

During this event, several DMSP satellites observed precipitating proton fluxes in the polar regions. We show here the DMSP/SSJ observations from two overpasses, sampling the cusp in the northern hemisphere and the nightside auroral oval in the southern hemisphere. Figure~\ref{fig:2011event_data} shows the comparison between simulated (Vlasiator) and observed (DMSP/SSJ) fluxes during those two overpasses. In Figure~\ref{fig:2011event_data}a and \ref{fig:2011event_data}b, the Vlasiator integrated energy flux of precipitating protons in the northern and southern hemispheres, respectively, are shown as a background colour plot in MLT--MLAT coordinates, in a similar way as in Figure~\ref{fig:dial2x2}. However, contrary to Figure~\ref{fig:dial2x2} where the data were shown at a fixed time step, the precipitating fluxes have here been time-averaged between $t = 900$ and $t = 1506$~s, smoothing out short-lived signatures associated with individual precipitation bursts. Overlaid on top of the time-averaged Vlasiator integrated energy fluxes are along-orbit observations from DMSP F18 and F16 sampling the dayside northern hemisphere and nightside southern hemisphere, respectively. It can be seen that the location of the cusp matches fairly well (Fig.~\ref{fig:2011event_data}a), the observations indicating a peak about one degree equatorwards from the Vlasiator peak, and centred around noon in both cases. The nightside auroral oval location is also in relatively good agreement between both data sets, albeit to a lesser extent (Fig.~\ref{fig:2011event_data}b).

To ease the comparison of simulated and observed fluxes in a quantitative manner, Figure~\ref{fig:2011event_data}c and \ref{fig:2011event_data}d show the corresponding time series of DMSP/SSJ measurements (orange lines) and that of the Vlasiator data along the orbit (green lines). For Vlasiator data, solid lines indicate the time-averaged integrated energy flux values along the orbit, the shadings give the $\pm 1 \sigma$ interval (with $\sigma$ the standard deviation), and the dotted lines give the maximum flux during $t=900$--1506~s. 

On the dayside (Fig.~\ref{fig:2011event_data}c), the DMSP integrated energy flux exhibits two peaks: the main one in the noon sector (at 07:20~UT), corresponding to the cusp signature and reaching \mbox{$5 \times 10^7$~keV\,cm$^{-2}$\,s$^{-1}$\,sr$^{-1}$}, and a secondary one in the pre-dusk sector (at 07:13~UT) reaching \mbox{$1.5 \times 10^7$~keV\,cm$^{-2}$\,s$^{-1}$\,sr$^{-1}$}. The Vlasiator integrated energy flux at corresponding locations exhibits a main peak in the cusp, reaching its maximum value of \mbox{2--$3 \times 10^7$~keV\,cm$^{-2}$\,s$^{-1}$\,sr$^{-1}$} (depending on whether one considers the time-averaged flux or the maximum flux) at a location about 1$^\circ$ more polewards in geomagnetic latitude. The cusp peak is therefore smaller by a factor of 1.7--2.5 in Vlasiator compared to DMSP, and it is also broader. In the pre-dusk sector, the Vlasiator flux forms a small secondary peak at about \mbox{$3 \times 10^6$~keV\,cm$^{-2}$\,s$^{-1}$\,sr$^{-1}$}, i.e. smaller by a factor of 5 than the corresponding peak in DMSP data. Between the two peaks, DMSP measured an extremely low proton flux, whereas Vlasiator fluxes are up to two orders of magnitude higher. This large discrepancy may come from the fact that the Vlasiator cusp signature is broader in MLT, or from the slight difference in the cusp peak geomagnetic latitude, or a combination of both. Therefore, while the agreement is poor where DMSP fluxes are low, the differences are not very large concerning the cusp peak flux values, considering that the range of measured fluctuations in integrated energy flux along the DMSP orbit spans over 2.5~decades (approximately a factor of 500 between the lowest and highest fluxes in observations).

The along-DMSP-orbit integrated energy flux on the nightside exhibits fewer spatial variations, DMSP observations essentially ranging within 3--$5 \times 10^7$~keV\,cm$^{-2}$\,s$^{-1}$\,sr$^{-1}$, while time-averaged Vlasiator fluxes are generally lower by a factor of 1--20 (Fig.~\ref{fig:2011event_data}d). The peak nightside DMSP flux is \mbox{$6 \times 10^7$~keV\,cm$^{-2}$\,s$^{-1}$\,sr$^{-1}$}, and the peak value of the maximum Vlasiator flux (green dotted line) is \mbox{$3 \times 10^7$~keV\,cm$^{-2}$\,s$^{-1}$\,sr$^{-1}$}, i.e. 50\% of the DMSP peak value. DMSP flux values fluctuate between \mbox{$2 \times 10^7$~keV\,cm$^{-2}$\,s$^{-1}$\,sr$^{-1}$} and \mbox{$6 \times 10^7$~keV\,cm$^{-2}$\,s$^{-1}$\,sr$^{-1}$} during 06:10–-06:13~UT, i.e. can vary by a factor of 3. The Vlasiator maximum flux values are comprised within those fluctuations in the DMSP data. While Vlasiator fluxes show an asymmetry between the pre-midnight and post-midnight sectors (with a peak centred around 23~MLT), this is not seen in DMSP observations. Finally, the fluctuations seen in DMSP data do not exist in Vlasiator fluxes; however, it is not possible to determine whether these fluctuations correspond to spatial or temporal features, given that the overpass took place over 9~min (comparable to the length of the analysed part of the Vlasiator run). It is important to keep in mind that the real-world IMF and solar wind conditions are neither uniform nor constant, hence a perfect match between observations and simulation cannot be expected.

The last two panels of Figure~\ref{fig:2011event_data} compare modelled and observed differential number fluxes of precipitating protons, again for the dayside (Fig.~\ref{fig:2011event_data}e) and nightside (Fig.~\ref{fig:2011event_data}f) overpasses. An immediately striking feature which can be identified is that the Vlasiator precipitation dataset misses the low-energy part of the precipitating proton spectra, since the lowest energy bin which is evaluated is at 500~eV (compared to 30~eV for DMSP SSJ observations). The isocontour plots, which correspond to the DMSP data, indicate that the lowest precipitating energies are often associated with high differential flux values (up to $\sim\! 3 \times 10^4$~proton\,cm$^{-2}$\,s$^{-1}$\,sr$^{-1}$\,eV$^{-1}$). The fact that those low energies are absent in the Vlasiator dataset can explain part of the discrepancies between Vlasiator and DMSP integrated energy fluxes (Figs~\ref{fig:2011event_data}c--d), especially when little precipitation is observed at higher proton energies (e.g. the secondary peak in the afternoon sector in Fig.~\ref{fig:2011event_data}c). However, in the energy domain where both Vlasiator and DMSP data are available, there is a reasonably good agreement between the two datasets regarding the differential number flux, in particular in the cusps (Fig.~\ref{fig:2011event_data}e within 0.5--10~keV) and in the nightside oval.

\begin{figure}
  \centering
  \includegraphics[width=0.95\columnwidth]{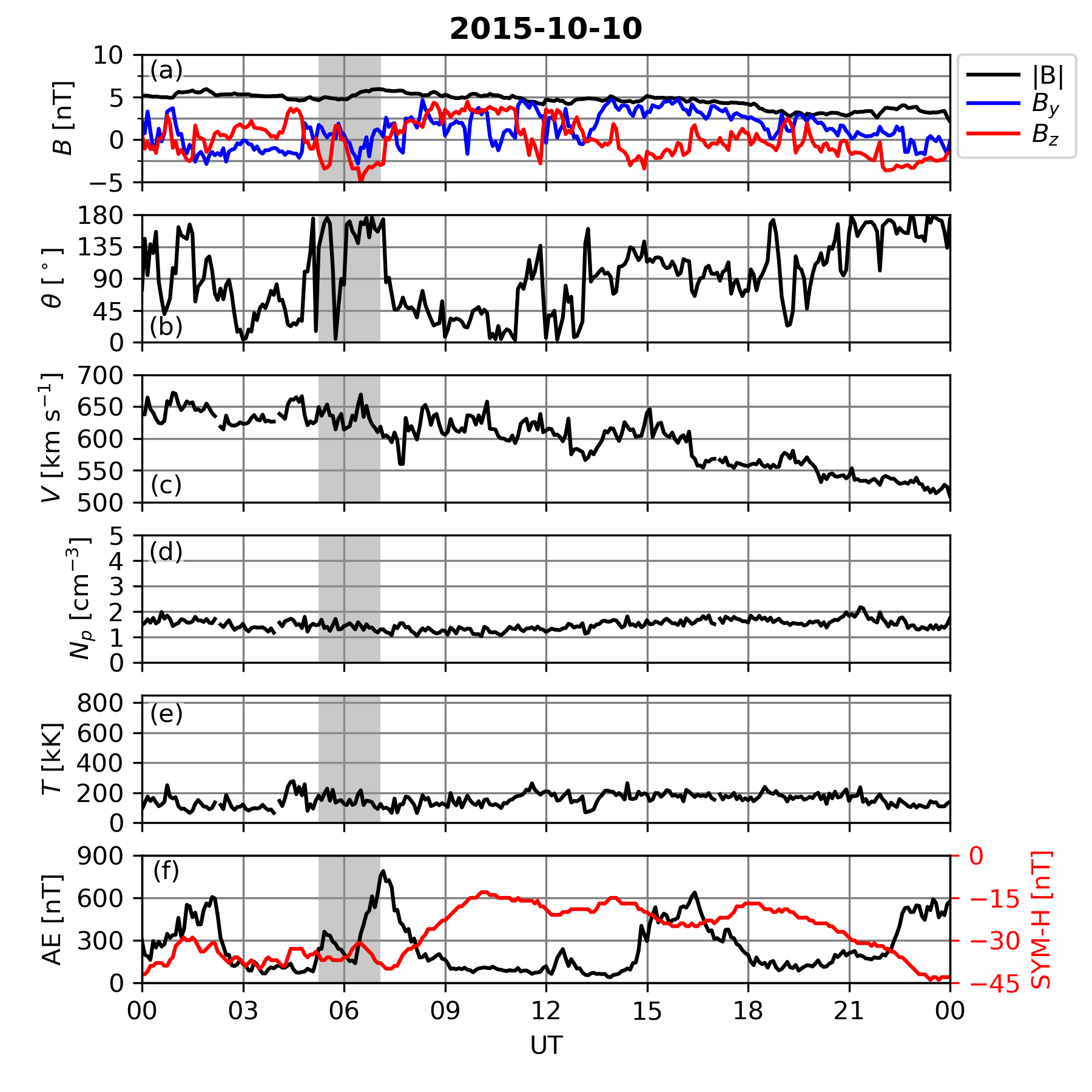}
    \caption{Solar wind conditions and geomagnetic activity on 10~October 2015. Same format as Figure~\ref{fig:2011event_omni}.}
    \label{fig:2015event_omni}
\end{figure}

While this first set of DMSP observations already exhibits similarities with the Vlasiator proton precipitation data, we also look at a second event under slightly different geomagnetic conditions, both to provide an additional baseline for comparison with Vlasiator fluxes and to discuss the variability in proton precipitation despite relatively similar solar wind and IMF conditions. Figure~\ref{fig:2015event_omni} presents the OMNI data corresponding to 10~October 2015, in the same format as Figure~\ref{fig:2011event_omni}. On that day, the driving conditions best resemble those used in the Vlasiator run within 5:30--7:00~UT (greyed area). The main differences compared to the earlier event are a lower solar wind temperature and greater geomagnetic activity, with the AE index peaking at $\sim$800~nT at 7:30~UT and SYM-H fluctuating between --30 and --40~nT (still on the same order as the value used in T01 for the mapping; see Sect.~\ref{sec:mapping}). 

\begin{figure}
  \centering
  \includegraphics[width=0.95\columnwidth]{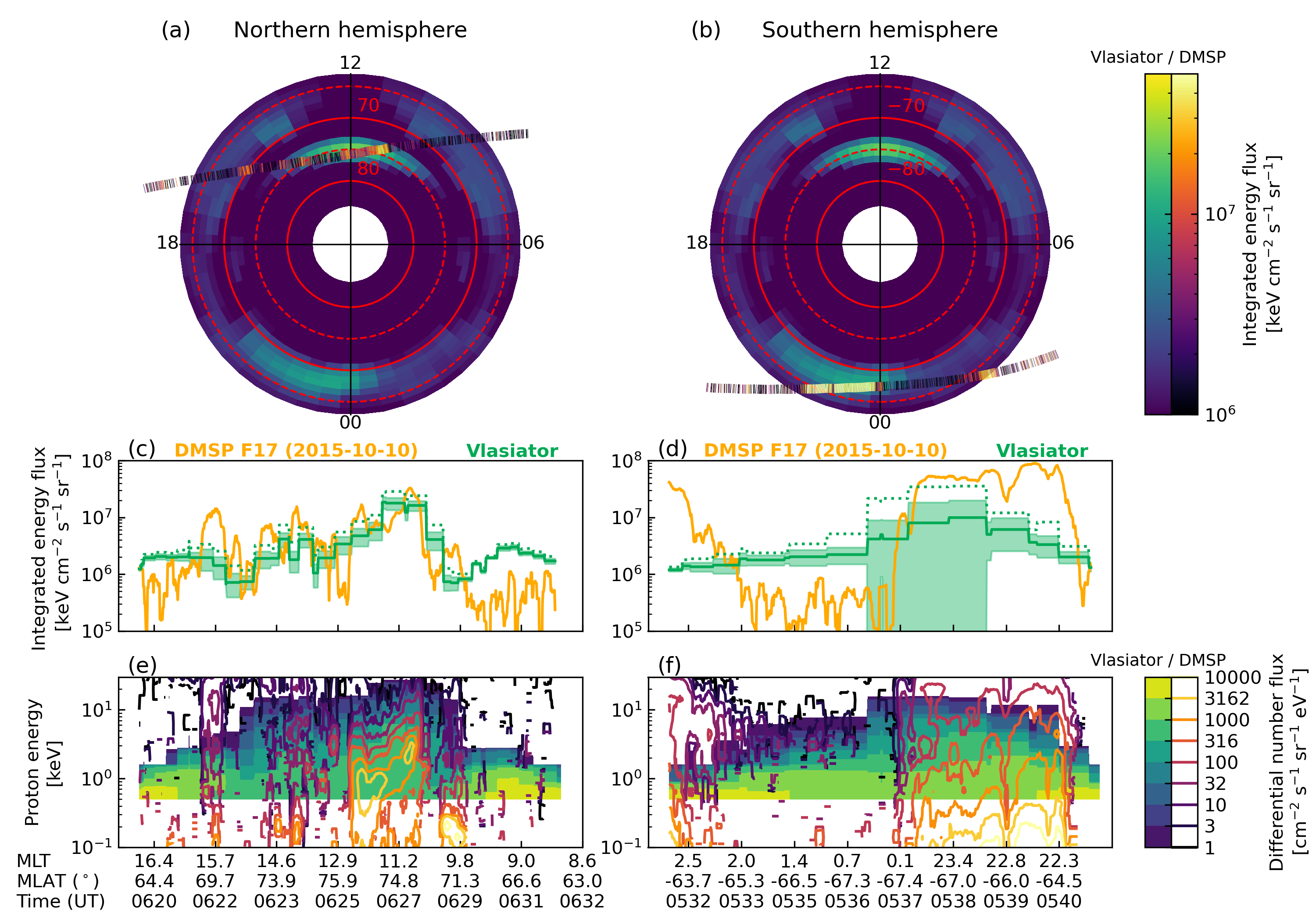}
    \caption{Comparison of Vlasiator proton precipitation and DMSP/SSJ observations on 10~October 2015. Same format as Figure~\ref{fig:2011event_data}, with overpasses from DMSP F17 in both hemispheres.}
    \label{fig:2015event_data}
\end{figure}

The comparison between Vlasiator and DMSP proton precipitation data is shown in Figure~\ref{fig:2015event_data}, in the same format as Figure~\ref{fig:2011event_data}, this time with DMSP data collected by the F17 spacecraft in both hemispheres. Here too, the top panels indicate a qualitatively reasonable match between the two datasets regarding the locations affected by proton precipitation. During this event, the cusp precipitation given by Vlasiator along the F17 satellite track is in better agreement with the observations than in the previous case, both in terms of integrated energy flux (Fig.~\ref{fig:2015event_data}c) and of differential number flux (Fig.~\ref{fig:2015event_data}e). The maximum fluxes within the cusp are almost exactly the same in both data sets (peak values around \mbox{2--$3 \times 10^7$~keV\,cm$^{-2}$\,s$^{-1}$\,sr$^{-1}$}), which is satisfactory given that the differences are significantly below the level of fluctuation seen in DMSP observations. The match is not as good for the nightside fluxes, as the integrated energy flux observed by DMSP/SSJ is either significantly lower or significantly larger than the time-averaged Vlasiator one, and the maximum Vlasiator flux reaches \mbox{$3 \times 10^7$~keV\,cm$^{-2}$\,s$^{-1}$\,sr$^{-1}$}, which is about 33\% of the highest value measured by the instrument \mbox{($9 \times 10^7$~keV\,cm$^{-2}$\,s$^{-1}$\,sr$^{-1}$, around 05:40~UT)}. Interestingly, the DMSP data exhibit a similar pre-midnight vs post-midnight asymmetry this time, like in Vlasiator (Fig.~\ref{fig:2015event_data}d and f), although the transition is very sharp in the satellite measurements. This asymmetry in the DMSP/SSJ data may either be due to an intensification of the nightside precipitation around 5:37~UT or due to a strong MLT dependence of the proton precipitation during this particular event. In the pre-midnight sector (after 5:37~UT), the differential number fluxes are in good agreement in the range where both datasets overlap (especially in the 0.5--3~keV energy range), suggesting that the difference in integrated energy fluxes can be largely explained by the absence of energies below 500~eV in the Vlasiator precipitation dataset. This can be investigated in a future run, for which the precipitation data reduction operator will be set to include protons of lower energies. We also note that observations contain protons at $\sim$10~keV energies in larger fluxes than the simulation.

Therefore, despite some inevitable differences, the shown DMSP observations suggest that the proton precipitation data produced during the studied Vlasiator run are realistic, in the sense that they can be found in nature during events when the solar wind and IMF driving conditions are relatively similar to those used in the simulation. Both events show that, even when the driving conditions (solar wind parameters and IMF) are similar, variability can be found in observations; obtaining simulated fluxes within the right order of magnitude is therefore very satisfactory.

\section{Discussion}
\label{sec:discussion}

\subsection{Comparison with other simulations and observations}
\label{sec:discussionOtherWorks}

We already found good agreement between the 2D--3V cusp precipitation patterns in \cite{Grandin2020} and coupled MHD and particle tracer simulations by \cite{Connor2015} and with 2.5D hybrid-PIC simulations by \cite{Omidi2007}. The 3D--3V results presented here exhibit similar characteristics in terms of energy--time dispersion and energy--latitude dispersion. Therefore, our results in the present study are also consistent with those past studies, with the added value that, for the first time, a self-consistent hybrid simulation of cusp proton precipitation was performed in 3D ordinary space, allowing the study of the MLT extent and variations of the polar cusp in both hemispheres. 

In our past 2D--3V study \citep{Grandin2020}, we showed that local differential number flux enhancements in the cusp were associated with the transit of flux transfer events, which led to bursts of proton precipitation at $< 30$~keV energies which exhibited energy dispersion as a function of time. The rate at which such bursts occurred in either cusp in 2D was found to be of about one burst per minute on average, compared to about 0.7~burst per minute in this 3D study. This slightly lower rate than in the 2D--3V run (which had otherwise identical driving conditions: $B_z = -5$~nT, $V = 750$~km\,s$^{-1}$, $N = 1$~cm$^{-3}$) can likely be explained by the fact that in the 2D--3V setup every solar wind magnetic field line needs to reconnect to the geomagnetic field at a rate imposed by the solar wind speed and IMF magnitude. The addition of a thin third spatial dimension to such a noon--midnight meridional plane run, in which the geomagnetic field had cylindrical symmetry along the $Y$ direction, led to $Y$-dependent reconnection at the dayside magnetopause \citep{PfauKempf2020}. Therefore, one can expect that a full 3D--3V run allows for more complex reconnection patterns and affect the formation rate of FTEs.

\cite{Zhang2013} carried out MHD simulations with the Lyon--Fedder--Mobarry global magnetospheric model, in which they investigated the cusp properties as a function of the solar wind driving and IMF conditions. For the simulated event, they found that the central MLAT of the cusp determined from the parallel ion number flux could be approximated with $\Lambda_\mathrm{ion} = 77.8^\circ + 0.78 B_z$ (for $B_z < 0$, in nanoteslas). For an IMF $B_z$ of $-5$~nT as in our Vlasiator simulation, this corresponds to a central cusp latitude of $73.9^\circ$, which is in very good agreement with Figure~\ref{fig:mlatkeo_day}b. This is also consistent with results from a statistical study by \cite{Zhou2000} from observations by the POLAR spacecraft, who also found that the centre MLAT of the cusp slightly shifts equatorwards at MLTs away from magnetic noon -- a slight trend consistent with this can be seen in Figure~\ref{fig:mlatkeo_day}a,c. Regarding the cusp centre in terms of MLT, multiple studies have shown a clear dependency on the IMF $B_y$ value \citep[e.g.][]{Karlson1996,Zhang2013,Zhou2000}. In our simulation with $B_y=0$, the cusp was found to be centred around noon (Fig.~\ref{fig:mltkeo_day}), which is consistent with previous studies. Finally, regarding the MLT width of the cusp, \cite{Maynard1997} estimated a low-boundary estimate of 3.7~hours in an observational event study, and \cite{Zhang2013} obtained values exhibiting a dependency on the solar wind dynamic pressure with a saturation around 6~hours, with a large spread between 0 and $\sim$4.5~hours for a dynamic pressure of 1.1~nPa like in our run. Again, looking at the isocontours for the 8.9~keV differential number flux precipitation in the cusp as a function of MLT (Fig.~\ref{fig:mltkeo_day}), our results are in good agreement with those findings.

On the nightside, we found that the energy-latitude dispersion of the proton fluxes in our simulation matches what \cite{Liang2014} refer to as ``reversed'' dispersion, with lower energies being found at lower latitudes. In their paper, they argued that a potential explanation for such a dispersion might be pitch-angle scattering of protons by H-band EMIC waves in the central plasma sheet, which can then lead to the precipitation of keV protons equatorwards from the high-energy ion isotropic boundary. It is beyond the scope of this paper to look in detail into the mechanism leading to this reversed dispersion in the present run, but it would prove interesting to investigate whether EMIC waves can indeed be found in the central plasma sheet in our run, as the spatial resolution in that region is 1000~km. A study by \cite{Dubart2020} found that EMIC waves could not grow in the magnetosheath in a (2D--3V) Vlasiator run with a spatial resolution of 900~km. Nevertheless, the plasma parameters in the tail being different, it might be possible that 1000~km suffices for H-band EMIC waves to grow in the central plasma sheet.

In a simulation with the RAM-SCBE model, \cite{Zhu2021} determined that beyond $L = 5$ the proton precipitation at keV energies is predominantly due to field line curvature in the magnetotail, whereas at lower $L$-shells scattering due to EMIC waves dominates. Since in our simulation the precipitating proton fluxes are evaluated at $r = 5.5\,R_E$, the main source for precipitation (besides injections from magnetotail reconnection sites) that we obtain on the nightside is therefore expected to be field line curvature due to the tail's topology. However, this mechanism should lead to a ``normal'' energy--latitude dispersion instead of a ``reversed'' one. This suggests that the reversed dispersion we obtain might simply be the signature of the reconnection-associated injections rather than of EMIC-wave scattering, as freshly reconnected magnetotail field lines dipolarise and low-energy protons, which travel slower along them compared to high-energy protons, hence end up at a lower latitude in the ionosphere.

While in future Vlasiator runs it will be desirable to push the inner boundary closer to the Earth and to investigate the contribution of EMIC waves to proton precipitation at lower $L$ values, it is worth noting that, according to \cite{Spanswick2017}, the bright proton aurora typically observed with ground-based optical instruments generally maps to $L=6$--10. This is consistent with the fact that the nightside proton oval is mostly resolved in the present run, despite the lowest MLAT values for which the mapping was possible being $\pm63.5^\circ$. Indeed, notwithstanding possible boundary effects, the proton fluxes clearly decrease before reaching those MLATs.

\subsection{Limitations of the current setup}
\label{sec:limitations}
While the results presented here are a significant improvement from the 2D--3V setup, there are still a few limitations which need to be discussed. First of all, the inner boundary of the Vlasiator simulation domain lies at 4.7~$R_E$ in this run, which sets a boundary on the lowest geomagnetic latitude for which the precipitating fluxes can be evaluated. Since the differential number fluxes have been mapped from 5.5~$R_E$ to avoid possible boundary effects, the most equatorward locations in the figures shown in this paper are correspondingly at a geomagnetic latitude of $\sim$63.5$^\circ$. During geomagnetically disturbed times, proton precipitation can frequently extend to lower latitudes \citep[e.g.][]{Galand2002}; therefore, to include the entire auroral oval, future simulations will need to have the inner boundary located more inwards. 

Still concerning the simulation setup, it is worth noting that, despite the spatial resolution being relatively coarse (2000~km close to the inner boundary in 3D--3V compared to 300~km in the 2D--3V simulations used in \citet{Grandin2019,Grandin2020}), the obtained precipitating proton fluxes are in remarkably good agreement with the 2D--3V results (proton energies and differential number fluxes are comparable to those obtained in the aforementioned studies) and in acceptable agreement with satellite observations during comparable driving conditions. While producing global simulations of near-Earth space at a better spatial resolution would be highly desirable, our results however suggest that most of the relevant kinetic effects leading to proton precipitation are reproduced in a satisfactory manner in this run. Yet, with such a resolution, we cannot study proton aurora structures at mesoscales \citep[10s to 100s~km;][]{Forsyth2020}, and the MLAT--MLT mapping consequently has a relatively coarse resolution (1$^\circ$ in MLAT, 0.5~h in MLT). It should nevertheless be pointed out that Vlasiator runs are extremely computationally expensive, as they consider a number of phase-space points on the order of $10^{12}$, which is on par with the number of particles included in modern cutting-edge PIC codes.

\begin{figure}
  \centering
  \includegraphics[width=0.8\columnwidth]{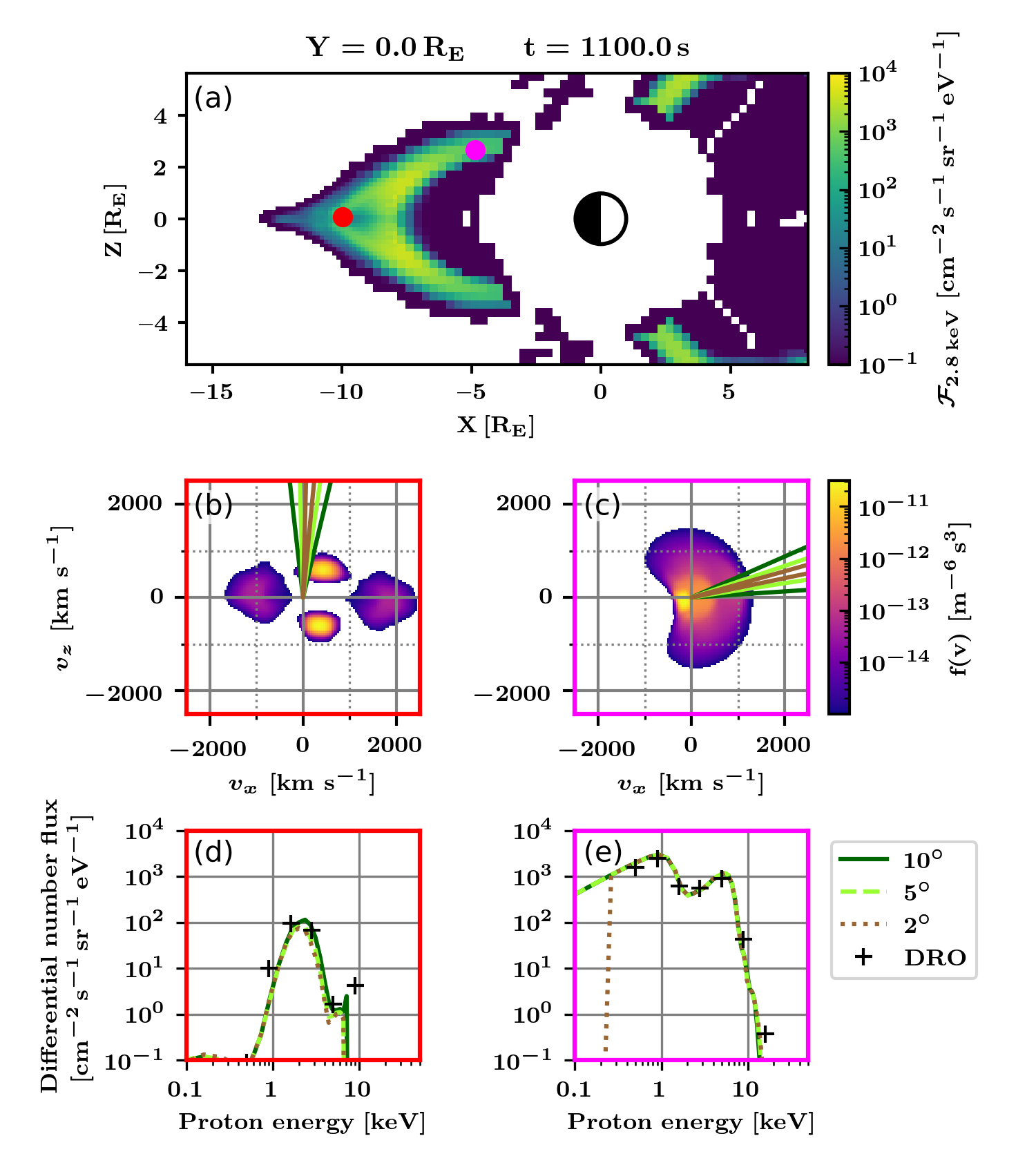}
    \caption{Comparison of differential number fluxes obtained with three loss-cone angle values. (a)~Differential number flux of precipitating protons at 2.8~keV obtained with the data reduction operator (DRO) at $t=1100$~s in the $Y=0$ plane of the simulation. The red (near $X = -10\,R_\mathrm{E}$) and magenta (near $X = -5\,R_\mathrm{E}$) dots indicate the locations at which velocity distribution functions (VDFs) were saved in the output file and will be analysed in the next panels. (b)~Slice in the ($v_x$,$v_z$) plane of the VDF at the location corresponding to the red dot. The boundaries of a loss cone of 10$^\circ$ (dark green), 5$^\circ$ (light green) and 2$^\circ$ (brown) are indicated. (c)~Same for the location corresponding to the magenta dot. (d)~Precipitating proton spectra at the red location obtained with the three loss cone values, with the same colour code as in the above panels. The `+' signs indicate the spectrum given by the DRO. (e)~Same for the magenta location.}
    \label{fig:losscone}
\end{figure}

On the methodology, three points can be discussed. First, the precipitating differential number fluxes are calculated everywhere assuming a loss-cone angle value of 10$^\circ$. While this choice is not too unreasonable close to the inner boundary, the bounce loss cone is actually a lot narrower in the magnetotail, having values as low as about 1$^\circ$ in the current sheet \citep{Sergeev1982}. Figure~\ref{fig:losscone} illustrates the effect of the choice for the loss cone angle value on the obtained precipitating differential number fluxes at two locations: one close to the current sheet (red dot in Fig.~\ref{fig:losscone}a), and one near the inner boundary (magenta dot). These two locations are chosen since they are among the few cells in which the full VDF is saved into the output file, enabling precipitating flux calculation a posteriori (in a similar manner as in the past 2D--3V studies). Three values are considered for the loss cone angle: 10$^\circ$, 5$^\circ$, and 2$^\circ$. The corresponding loss cones are shown in velocity space in Figures~\ref{fig:losscone}b,c, and the resulting differential number fluxes are given in Figures~\ref{fig:losscone}d,e. We can see that there are almost no noticeable differences in the three precipitating proton spectra near the inner boundary, and the discrepancies remain minor even in the current sheet. The narrowest loss cone angle (2$^\circ$) does not enable the calculation of fluxes below $\sim$300~eV (see Fig.~\ref{fig:losscone}e), as the lower-energy boundary is constrained by the velocity-space resolution, for a given loss-cone angle. The black `+' signs indicate the differential flux values obtained by using the data reduction operator (DRO) calculating the fluxes in 9~energy bins while Vlasiator is being run. Small differences arise due to the averaging over larger energy bins than when analysing a posteriori the proton VDF, which can be done on a fine energy grid since the computational load is not an issue when working on an output file. Nevertheless, the overall shape of the spectrum is well captured by the DRO. Therefore, the advantage of getting precipitation information in every cell thanks to the DRO largely supersedes the high energy resolution on the obtained fluxes at the very few locations where the proton VDF is saved in the output files. In future runs, we might also consider calculating the differential fluxes for more than 9 energy bins to increase the energy resolution and capture lower proton energies.

Second, the symmetrical patterns of proton precipitation between the northern and southern hemispheres obtained in this study are largely due to the simulation setup chosen for this Vlasiator run. Indeed, with a purely dipolar geomagnetic field aligned with the $Z_\mathrm{GSE}$ axis, geomagnetic coordinates are exactly superposed to geographic ones. Therefore, interhemispheric asymmetries arising from the non-dipolar and non-tilted nature of the geomagnetic field cannot be reproduced in the simulation. More importantly, the fact that the simulation is driven with purely southward IMF conditions implies that effects associated to the IMF $B_y$ are not included. It is however clear that, in nature, the IMF seldom has a purely southward orientation and varies rapidly with time (as can be seen in Figs~\ref{fig:2011event_omni}a and \ref{fig:2015event_omni}a). It has been shown that the $B_y$ component of the IMF can lead to interhemispheric asymmetries in the intensity of the aurora not only on the dayside \citep{Qiu2022}, but also on the nightside \citep{Liou2019}. Therefore, future work could explore to what extent these asymmetries can be found in auroral proton precipitation obtained with Vlasiator, should a 3D--3V run with a nonzero IMF $B_y$ become available.

Third (and finally), the effects of the ionospheric feedback on the magnetosphere are not modelled. While it is generally considered that protons of multi-keV energies are not highly affected by field-aligned potential drops above the ionosphere \citep{Liang2013}, they might have some relevance in the lower-energy component of auroral protons. More generally, due to the very crude description of the ionosphere in the current version of Vlasiator (where the inner boundary is a perfectly conducting sphere), the absence of a realistic ionospheric feedback on the magnetosphere might explain for instance the discrepancies between Vlasiator fluxes and DMSP observations on the nightside. In reality, the ionospheric conductivities and horizontal currents are complex and can exhibit small-scale ($<10$~km) structuring \citep[e.g.][]{Juusola2016,Lysak2002,Streltsov2004} as well as asymmetries between the northern and southern hemisphere, which affects the field-aligned currents coupling the ionosphere to the magnetosphere, especially in the nightside \citep[e.g.][]{Workayehu2021}. As a consequence, nightside auroral precipitation patterns generally exhibit complex spatiotemporal variability as seen in the DMSP observations (Figs~\ref{fig:2011event_data} and \ref{fig:2015event_data}) and which are not reproduced in Vlasiator, partly explaining orders-of-magnitude discrepancies in some portions of the orbit in the model--observations comparisons. Therefore, including a more realistic coupling between the inner boundary and the ionosphere would be a highly valuable development for future precipitation studies, especially when discussing nightside auroral precipitation.

\section{Conclusions}
\label{sec:conclusions}

We presented here the first results on auroral proton precipitation in a global 3D--3V hybrid-Vlasov simulation of near-Earth space with constant, southward IMF driving conditions. The main results can be summarised as follows:
\begin{enumerate}
    \item On the dayside, proton precipitation in the polar cusps occurs asynchronously between both hemispheres and takes place in the form of bursts at an average rate of 0.7~min$^{-1}$. These bursts are correlated with the transit of flux transfer events in the high-altitude cusp.
    \item The cusp morphology (centre latitude, MLT, width) is in good agreement with past observational and modelling (MHD) studies, given the solar wind and IMF driving parameters.
    \item On the nightside, the simulation clearly shows that proton precipitation can be quite localised in MLT depending on where the reconnection takes place in the magnetotail; correlation has been found with the presence of fast Earthward flows in the near-Earth magnetotail. 
    \item Nightside fluxes exhibit symmetry between the two hemispheres, but a ``reversed'' energy--latitude dispersion is found, which is in contradiction with the expected pattern generated by the field line curvature mechanism for loss-cone scattering of keV protons and is likely simply due to the injection of protons through magnetotail reconnection.
    \item The peak values of simulated precipitating fluxes are within an acceptable range of discrepancy when compared with particle observations from DMSP/SSJ during similar driving conditions, both qualitatively and quantitatively. The agreement is better for the cusp precipitation than for the nightside oval precipitation, presumably because of the absence of ionospheric feedback in the simulation and the idealised simulation setup which departs from the real-world driving conditions during the selected events. Due to the localised nature of cusp precipitation leading to large spatial gradients in integrated energy flux of proton precipitation, a reasonable ($\sim$1$^\circ$) difference in the geomagnetic latitude of the precipitation peak can result in poor agreement in the off-noon portions of the DMSP satellite orbits.
\end{enumerate}

Future work will include making use of this method to estimate the total energy input during an individual substorm associated to proton precipitation using Vlasiator simulations and observations. Another avenue will be investigating whether EMIC waves can grow in the simulation despite the relatively coarse spatial resolution, and what role they might play in the auroral proton precipitation. Finally, simulations with other sets of input parameters and a better description of the feedback from the ionosphere will enable the analysis of their effects on auroral proton precipitation patterns. Such endeavours will contribute to improving the understanding and the prediction of energy input into the upper atmosphere via particle precipitation, which is paramount to refine space weather forecasting.

\begin{acknowledgements}
We acknowledge the European Research Council for starting grant 200141-QuESpace, with which the Vlasiator model was developed, and consolidator grant 682068-PRESTISSIMO awarded for further development of Vlasiator and its use in scientific investigations. We gratefully acknowledge Academy of Finland grant numbers 338629-AERGELC'H, 336805-FORESAIL, 322544-UNTWINE, 339756-KIMCHI, and 335554-ICT-SUNVAC. The Academy of Finland also supported this work through the PROFI4 grant (grant number 3189131). The work of HZ is supported by the University of Helsinki (three-year research grant  2020--2022 -- RESSAC) The CSC -- IT Center for Science and the PRACE Tier-0 supercomputer infrastructure in HLRS Stuttgart (grant number 2019204998) are acknowledged as they made these results possible. The authors wish to thank the Finnish Grid and Cloud Infrastructure (FGCI) for supporting this project with computational and data storage resources.

We gratefully acknowledge the DMSP/SSJ community and data stewards and services for providing observations of precipitating particle fluxes, which were retrieved from \url{http://cedar.openmadrigal.org/}. We also thank J.H. King and N.E. Papitashvili of NASA/SPDF and QSS and the original providers of the input ACE, Wind and IMP 8 data for making the OMNI data available. OMNI data were retrieved from \url{https://omniweb.gsfc.nasa.gov/}, at 5-min resolution \citep{OMNI_data_5min}.
\end{acknowledgements}


\bibliography{biblio}

\end{document}